\documentclass[iop]{emulateapj}

\usepackage{color}
\usepackage{color}
\usepackage{rotating}
\usepackage{lscape}

\usepackage{multirow} 

\newcommand{\qsoname}  {SDSS J104257.58+074850.5}
\newcommand{\galaxyname} {GQ1042+0747}
\newcommand{\HI}{\ion{H}{1}~}
\newcommand{\kms}{km~s$^{-1}$ }
\newcommand{\Lya}{Lyman~$\alpha$~}

\shorttitle{GBT Absorption Survey for \HI\ in Halos of Galaxies}
\shortauthors{Borthakur et al.}

\begin{document}

\title{A GBT Survey for \HI\ 21~cm Absorption in the Disks and Halos of Low-Redshift Galaxies \altaffilmark{1}}

\altaffiltext{1}{Based on observations with (1) the telescopes of the
National Radio Astronomy Observatory, a facility of the National
Science Foundation operated under cooperative agreement by Associated
Universities, Inc., and (2) the Apache Point Observatory 3.5m telescope, which is owned and
operated by the Astrophysical Research Consortium.}

\author{Sanchayeeta Borthakur, Todd M. Tripp, Min S. Yun}
\affil{Department of Astronomy, University of Massachusetts, Amherst, MA 01003, USA}
\email{sanch@astro.umass.edu}

\author{David V. Bowen }
\affil{Princeton University Observatory, Peyton Hall, Ivy Lane, Princeton NJ 08544}

\author{Joseph D. Meiring}
\affil{Department of Astronomy, University of Massachusetts, Amherst, MA 01003, USA}

\author{Donald G. York }
\affil{Department of Astronomy and Astrophysics, University of Chicago, Chicago, IL 60637, USA; Enrico Fermi Institute, University of Chicago, Chicago, IL 60637, USA}

\author{Emmanuel Momjian}
\affil{National Radio Astronomy Observatory, 1003 Lopezville Road, Socorro, NM 87801, USA}


\begin{abstract}
We present a \HI\ 21~cm absorption survey with the {\it Green Bank Telescope} (GBT) of galaxy-quasar pairs selected by combining galaxy data from the {\it Sloan Digital Sky Survey} (SDSS) and radio sources from the {\it Faint Images of the Radio Sky at Twenty-Centimeters} (FIRST) survey. Our sample consists of 23 sightlines through 15 low-redshift foreground galaxy - background quasar pairs  with impact parameters ranging from  1.7~kpc up to 86.7~kpc. We detected one absorber in the GBT survey from the foreground dwarf galaxy, \galaxyname, at an impact parameter of 1.7~kpc and another possible absorber in 
our follow-up Very Large Array (VLA) imaging of the nearby foreground galaxy UGC~7408. Both of the absorbers are narrow (FWHM of 3.6 and 4.8~\kms), have sub Damped \Lya column densities, and most likely originate in the disk gas of the foreground galaxies. We also detected \HI\ emission from three foreground galaxies including UGC~7408. Although our sample contains both blue and red galaxies, the two \HI\ absorbers as well as the \HI\ emissions are associated with blue galaxies. We discuss the physical conditions in the 21~cm absorbers and some drawbacks of the large GBT beam for this type of survey.

\end{abstract}

\section{INTRODUCTION\label{intro}}

Inflow and outflow of gas between the intergalactic medium (IGM) and the disks of galaxies is a crucial aspect of galaxy evolution. In the last decade our theoretical understanding on how galaxies acquire gas has advanced considerably. 
Traditionally, galaxies were believed to accrete via the ``hot" mode \citep{white_ress78,white_frank_91}, where accreting gas shock-heats to the virial temperature and then all halo gas within a ``cooling radius'' cools to form stars. Recent hydrodynamic cosmological simulations have discovered that gas accretion also occurs in the so called ``cold" mode where the gas is able to cool as it descends into galactic potentials and thus never approaches the virial temperature \citep[see ][]{birn03,keres05}.
Unlike the hot-mode, the cold-mode accretion occurs through long filaments connecting the extended disks and halos of galaxies to their surrounding intergalactic medium (IGM).
Although initial results by \citet{keres05} showed that the cold mode accretion is dominant in galaxies with lower masses ($< 10^{10.3}~\rm M_{\odot}$) at redshift $z=0$, recent simulations by \citet{keres09} have shown that cold-mode accretion also persists even in massive galaxies with hot envelopes. 
In addition to filamentary accretion of cool gas from the intergalactic medium, cool gas is also expected to condense out of the hot galactic halos. \citet{mall_bull04} argued that small clouds of cool gas condense out of hot gaseous halos through a multiphase process due to thermal instabilities. These halo clouds are predicted to have a characteristic size of $\approx$ 1 kpc and are expected to be another continuous source of cool gas for galaxies.

On the other hand, a substantial amount of material is ejected from galactic disks into their halos via galactic winds \citep[see the review by][]{veilleux05}. Evidence of outflows has been detected in low-ionization metal absorption lines of Na I \citep[e.g.][]{schwartz04,rupke05a,rupke05b} and Mg II \citep{tremonti07,rubin10}. 
 \citet{rubin10} found that the strength of the outflow correlates with the absolute star-formation rate (SFR) and claimed that the mass outflow rates were of the order of the SFRs for their sample.  Assuming a negligible halo drag, \citet{rupke05b} found that only 5-10\% of the neutral material in starburst-driven winds escape into the IGM. In other words, the wind speeds are not larger than the galactic escape velocity for most galaxies and thus the outflowing material is primarily retained in the halos of galaxies.  According to the Galactic Fountain model \citep[]{shapiro76}, the baryonic material deposited by the winds can eventually cool down and rain back into the galaxies. This model and subsequent work by \citet{bregman80} predicts the existence of high-velocity clouds (HVCs) as a by-product of the cooling of the hot gas. 

Besides inflow and outflow signatures, there is growing evidence of the extended nature of the galactic disks. \ion{Mg}{2} absorbers, which trace cool low-ionization gas, have been believed to be associated with extended disks \citep{wagoner67,bowen95,charlton96}. Observations of sightlines probing foreground galaxies at impact parameters up to 75~$h^{-1}$~kpc  with inclination angles between 40$^{\circ}$ to 75$^{\circ}$ have confirmed \ion{Mg}{2} absorbers exhibiting disk-like rotation \citep{steidel02,kacprzak10}.
 \citet{steidel02} found that four of five background QSOs that lie to one side of a foreground galaxy exhibit \ion{Mg}{2} absorption from the latter that are  shifted in the same sense from the systemic redshift of the foreground galaxy. They concluded that the kinematics of their sample could be understood in terms of a disk-like rotation combined with a lagging halo. This was confirmed in a recent study by \citet{kacprzak10}, where most of the \ion{Mg}{2} absorption systems in their sample fully reside to one side of the galaxy systemic velocity. They also found a close coupling between galaxy inclination and \ion{Mg}{2} absorption velocity spread.  While it is clear that there is cool gas at these distances (~$>$~100~kpc for $h = \rm 0.7 ~{\rm km~s}^{-1}~{\rm Mpc}^{-1}$), it is unclear how much baryonic mass exists in these systems. Part of the problem is that these observations do not probe the hydrogen directly, and the coupling of neutral gas and metals  is not well understood.

To properly characterize the total baryonic content in the extended disks as well as gaseous halos, \HI\ observations must be made at a wide range of distances from the galaxies. 
Absorption spectroscopy is a powerful tool for probing low-column density systems with high sensitivity independent of the redshift of the absorbing material.
 Unfortunately, the \Lya transition for $z<~1.7$ galaxies occurs in the ultraviolet (UV) band and thus ground-based observations of these systems are not possible. 
 However, the \HI\ 21~cm hyperfine transition for low$-z$ systems can be observed with existing radio facilities. Various authors \citep[e.g.,][]{kanekar97,lane01,vermeulen03, darling04, keeney05, gupta07, gupta09, gupta10} have probed neutral gas in extragalactic sources at various redshifts using 21~cm absorption. 
 In their recent study of the radio galaxy Centaurus~A, \citet{struve10} discovered an \HI\ 21~cm absorbing cloud of column density $\sim \rm 10^{20}~\rm cm^{-2}$ (assuming T$_s = 100~K$ and filling factor 1) in a region beyond the optical extent of the galaxy, at a distance of 5.2~kpc from the galactic nucleus using the northeastern radio lobe of the galaxy itself as a background source.
 While there is a substantial amount of literature on the using \HI\ absorption toward radio lobes to probe the immediate vicinity of AGN \citep[e.g.][]{saikia09}, this example shows how radio galaxies can also be used to probe the ISM of galaxies far from the nuclear region."

Another advantage of radio absorption studies is that radio sources are often extended and thus unlike UV or optical absorption spectroscopy, which is a single pencil beam study, absorption in the radio band provides a unique opportunity to probe multiple sight lines through the ISM of  foreground galaxies simultaneously, and thus is an excellent way to map the small-scale structures in extragalactic ISMs. Such small-scale structures down to AU-scales were detected in the Milky Way as early as in 1976 by \citet{dieter76} and later confirmed by \citet{diamond89}. 
Over the last few years, multi-epoch mapping of the extended radio emission of quasars by \citet{brogan05} and \citet{lazio09} have confirmed the small-scale structures in the ISM of the Milky Way detected via 21~cm absorption.
These authors report that the covering fraction of these structures are roughly 10\% which implies that their volume filling fraction is only about 1\%. 
Similar spatial variations in the ISM of galaxies outside the Milky Way have not yet been detected.
One way to constrain the sizes and physical nature of the cold clouds, including their spin temperature and covering fraction, is to conduct a survey of \HI\ absorption against multiple bright background sources that probe a single galaxy. 
Such a survey would provide unique information on the (1) structure of the cold gas clouds in a galaxy, (2) variation of cloud properties as a function of distance from the galaxy/galactic disk; and (3) the cold gas distribution as a function of global properties (such as color, stellar mass) of galaxies.

In this paper, we investigate the existence and properties of cold gas in the disks and halos of low-redshift galaxies ($z<0.3$). We seek to constrain the frequency and strength of the \HI\ distribution as function of the distance from the center of a galaxy. A similar study was carried out by \citet{gupta10} with 5 quasar-galaxy pairs using the {\it Giant Metre Radio Telescope} (GMRT). The most important difference between their observational set up and ours is in the spectral and spatial resolution. The two studies are  highly complementary. On one hand, the study conducted by \citeauthor{gupta10} has a high spatial resolution which enables them to detect absorption more efficiently.  On the other hand, our study achieved a high spectral resolution, which is crucial in identifying narrow \HI\ features. The 21~cm absorbing gas is expected to be cold and thus the line widths can be extremely narrow. Such features were observed in the Milky Way \citep[$\sim$1~$\rm km~s^{-1}$ by ][]{knapp72}, the Large Magellanic Cloud (LMC) \citep[1.4~$\rm km~s^{-1}$ by][]{dickey94}, and Small Magellanic Cloud (SMC) \citep[1.2~$\rm km~s^{-1}$ by][]{dickey00}. Unlike the extended sample used by \citet{gupta10} in their analysis, we use an unbiased sample, which was selected solely on the basis of the impact parameter without any consideration to absorption detection through other line transitions.

 \begin{figure*}[t]
 \figurenum{1}
\begin{tabular}{l l l }
\includegraphics[scale=0.275,angle=-0]{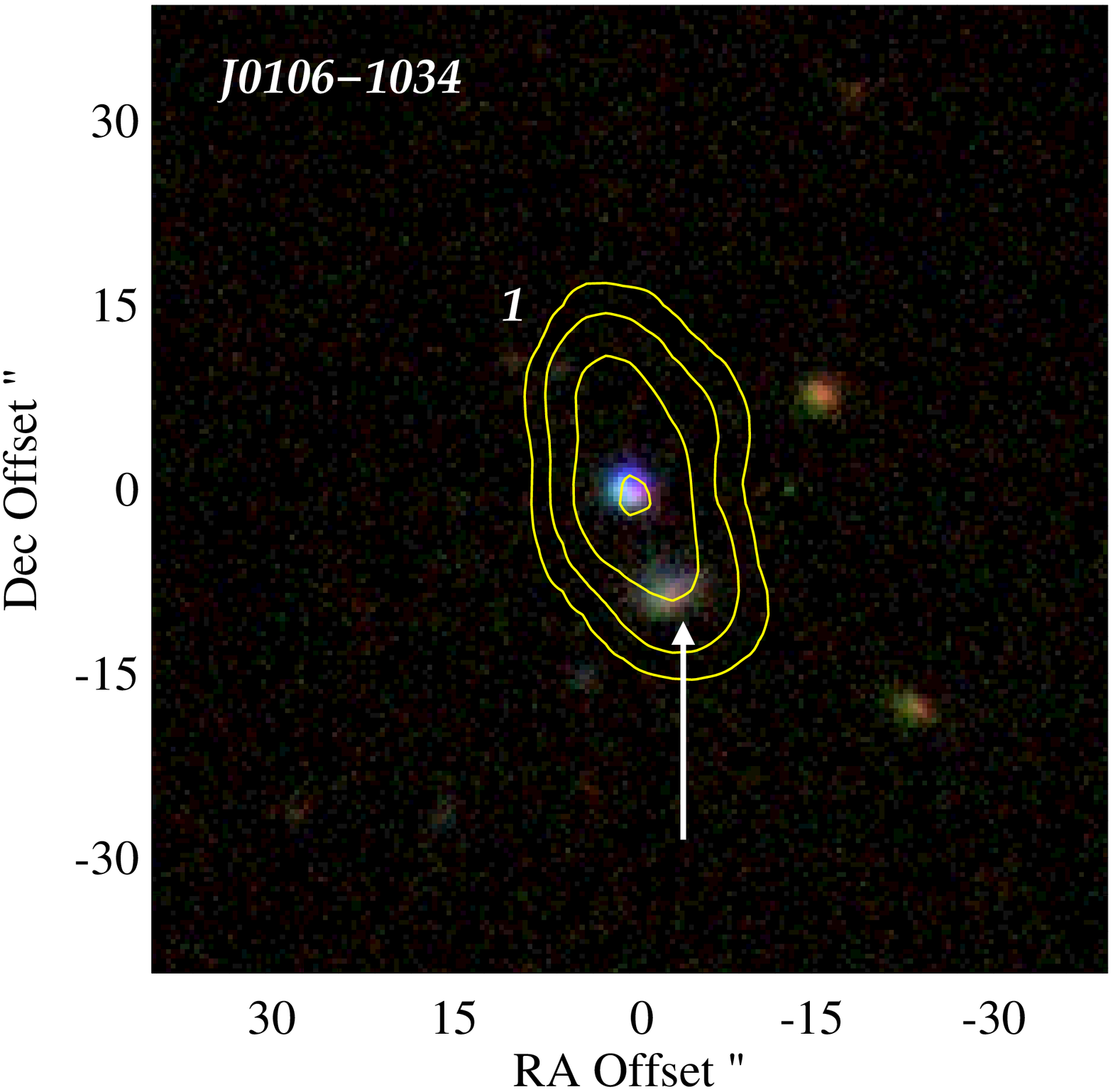} & \includegraphics[scale=0.275,angle=-0]{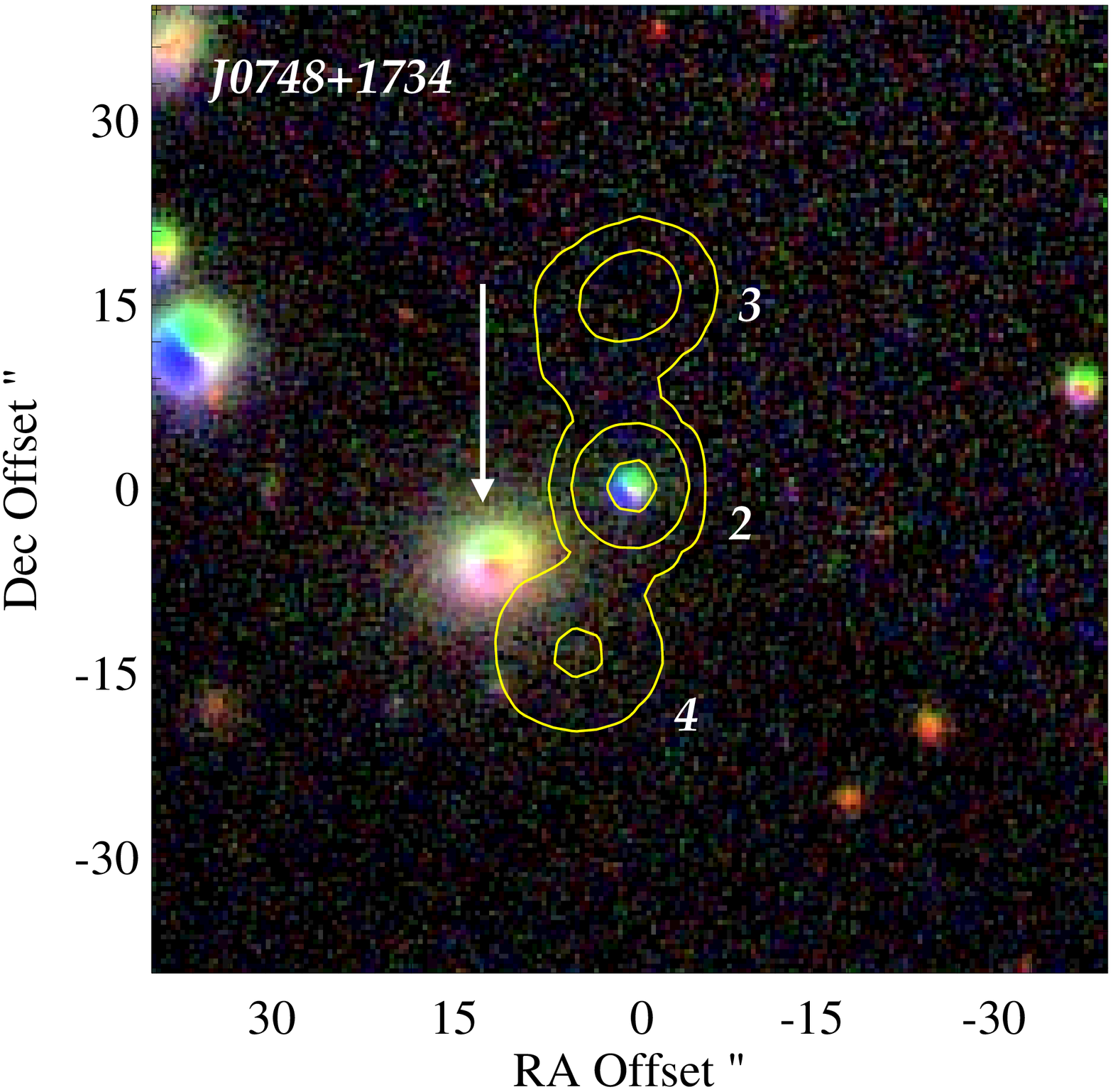} & \includegraphics[scale=0.275,angle=-0]{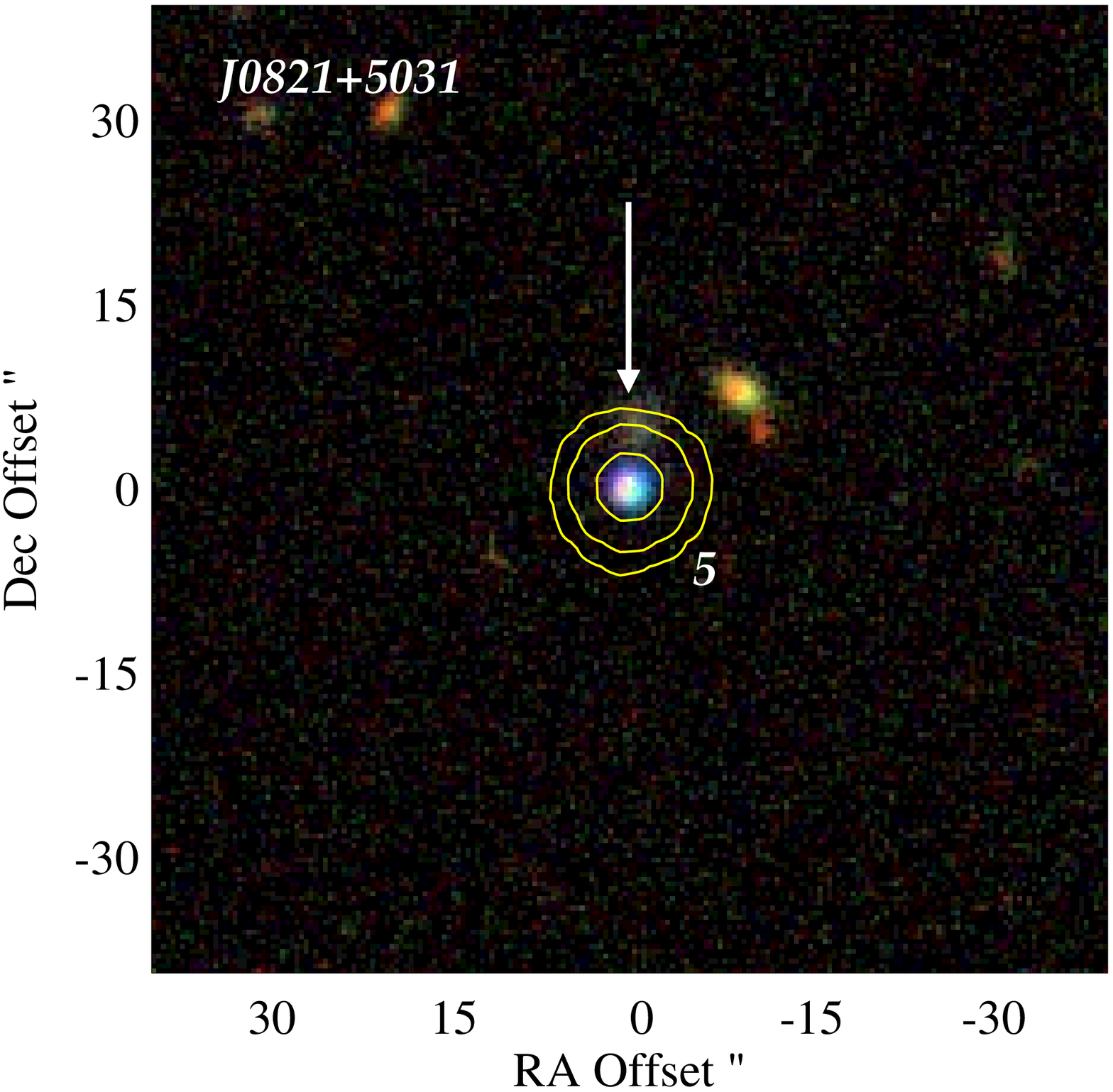}  \\
\includegraphics[scale=0.275,angle=-0]{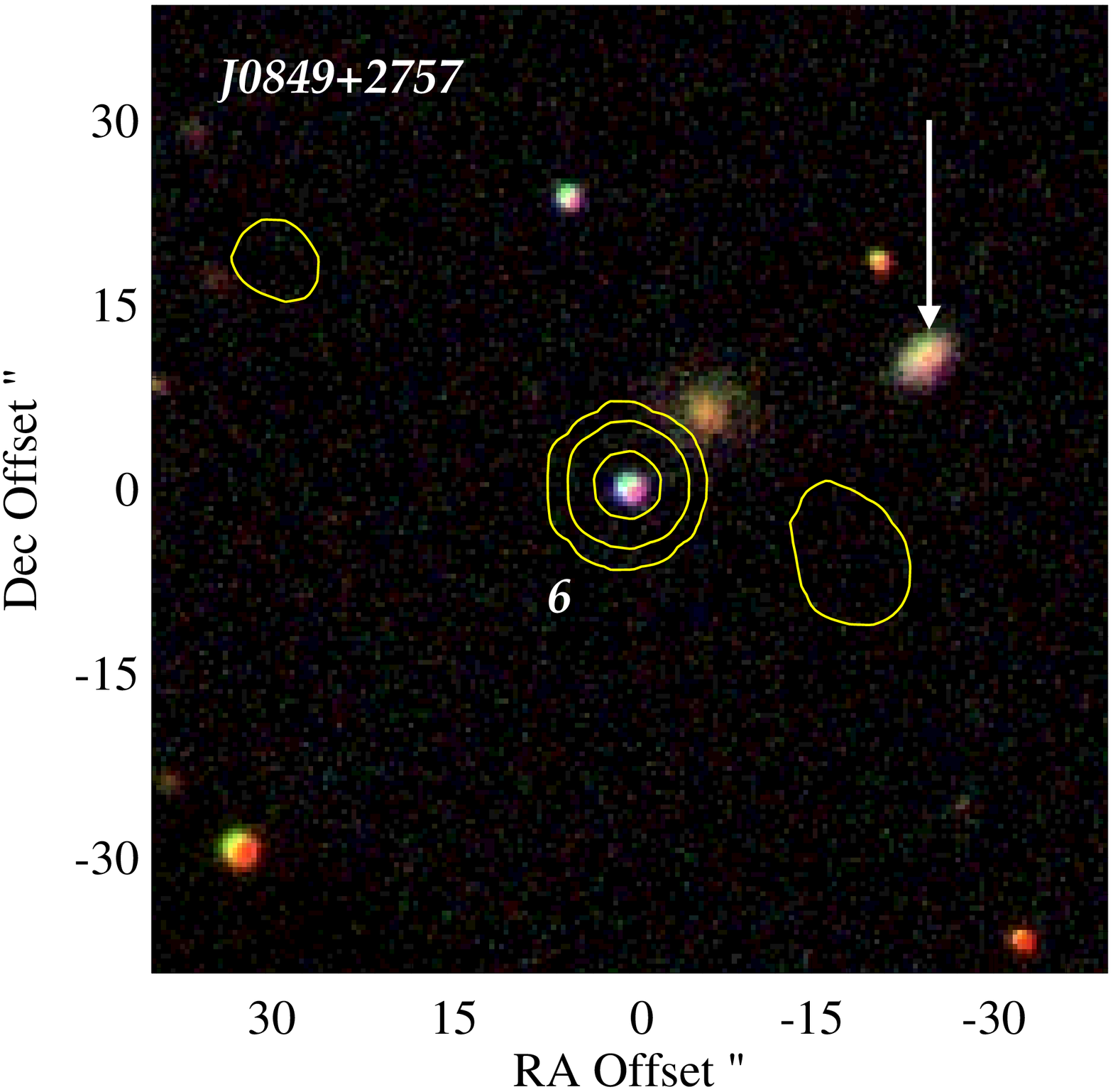} & \includegraphics[scale=0.275,angle=-0]{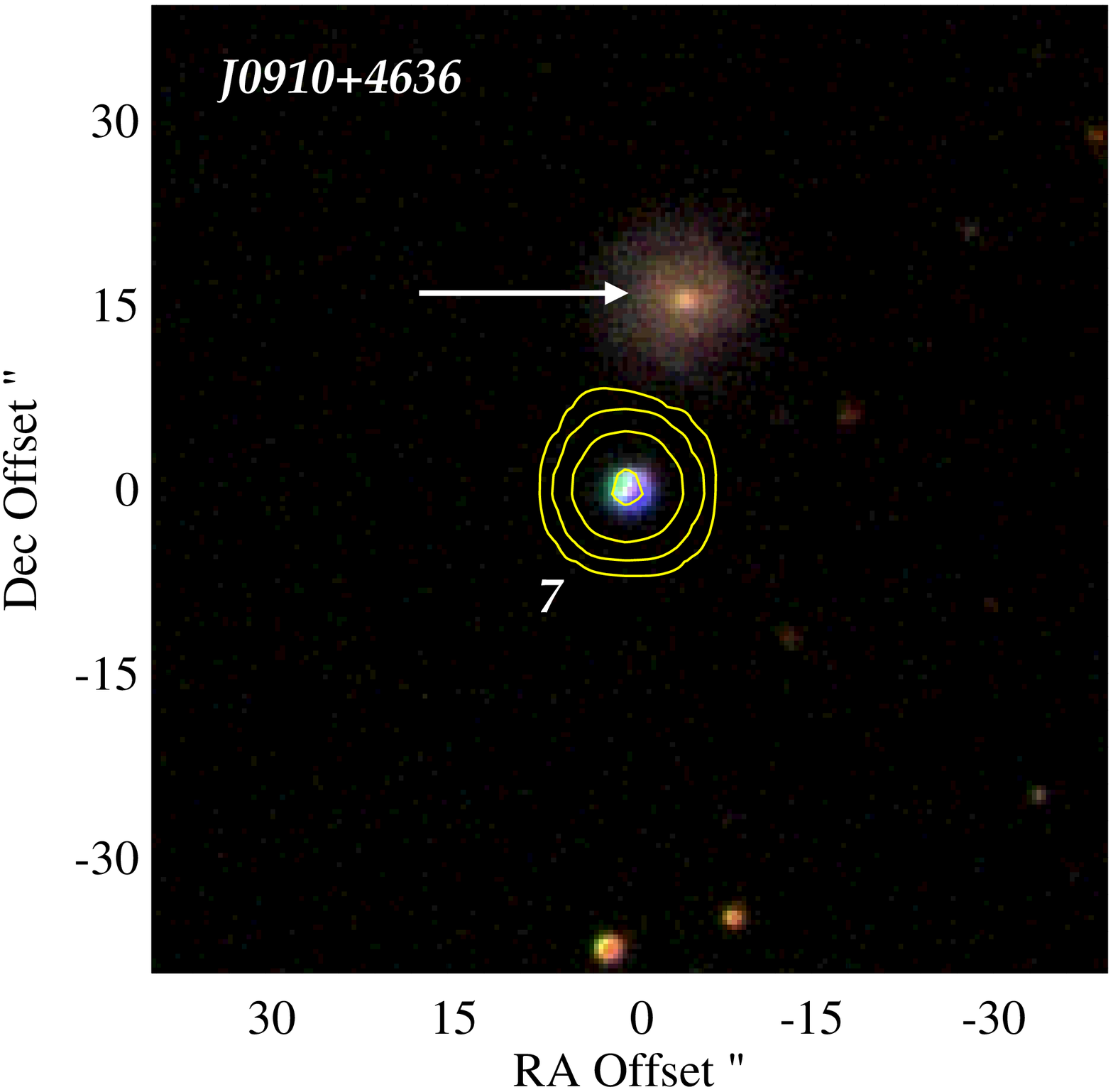} & \includegraphics[scale=0.275,angle=-0]{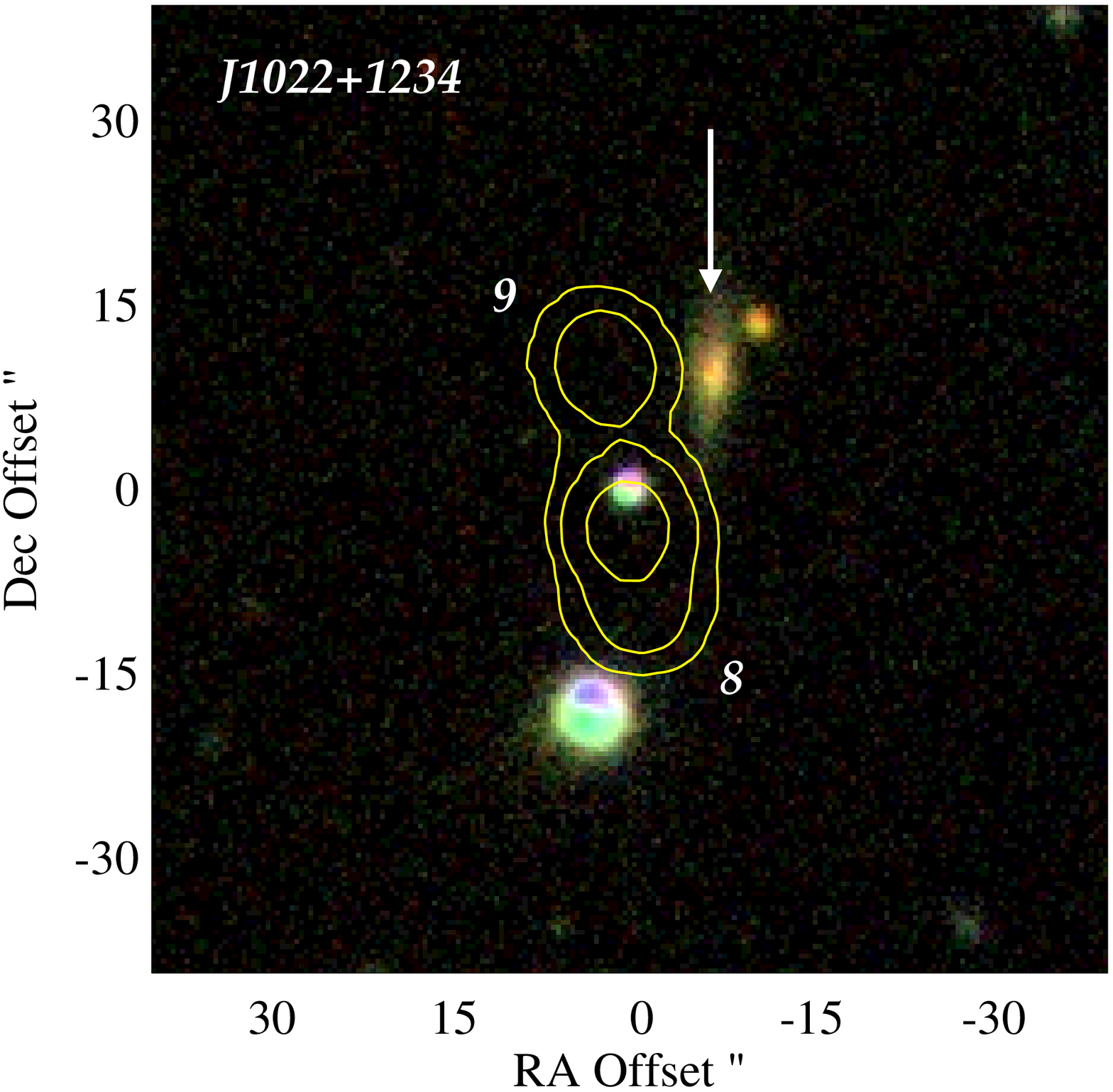} \\
\includegraphics[scale=0.275,angle=-0]{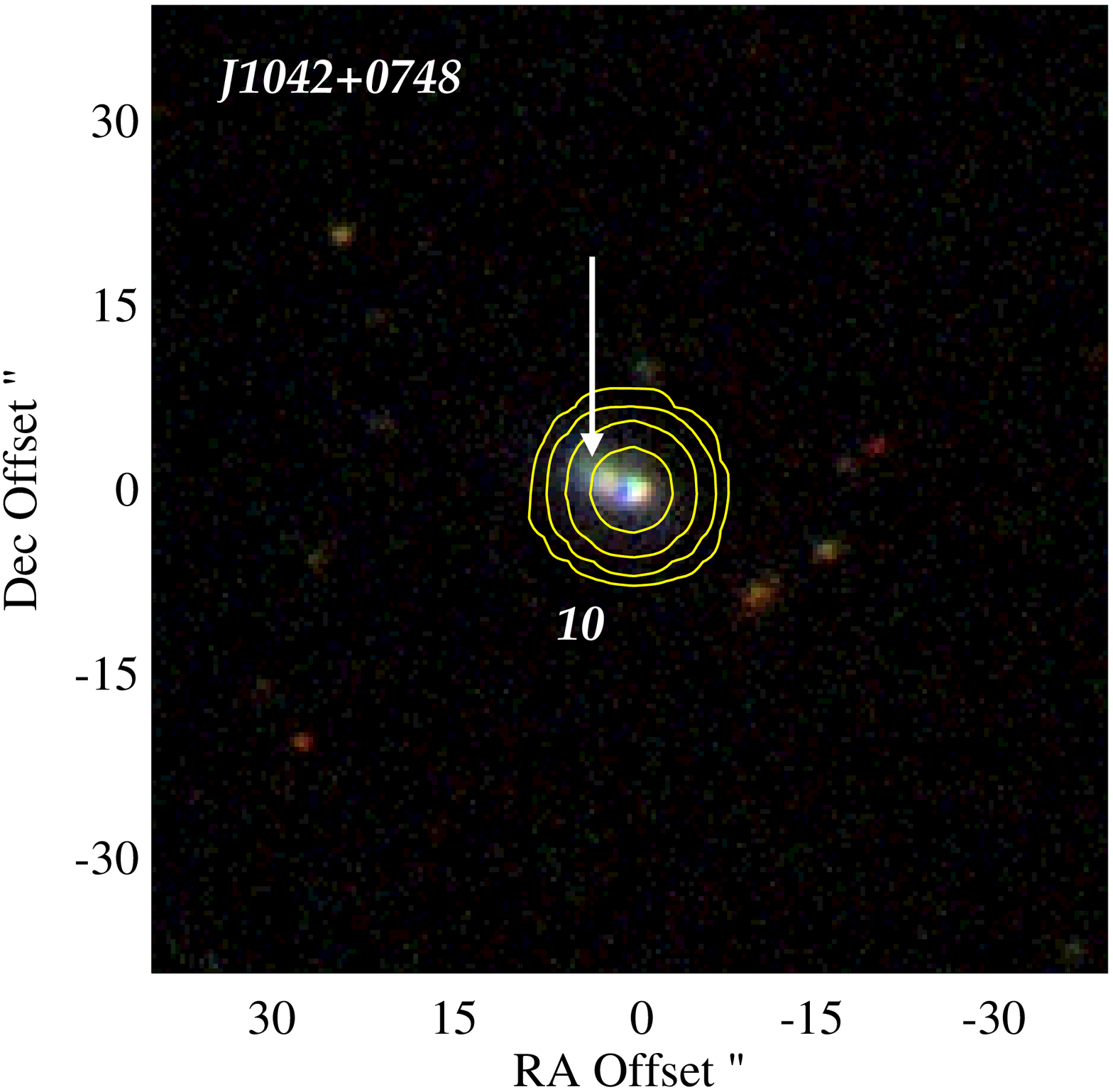} & \includegraphics[scale=0.275,angle=-0]{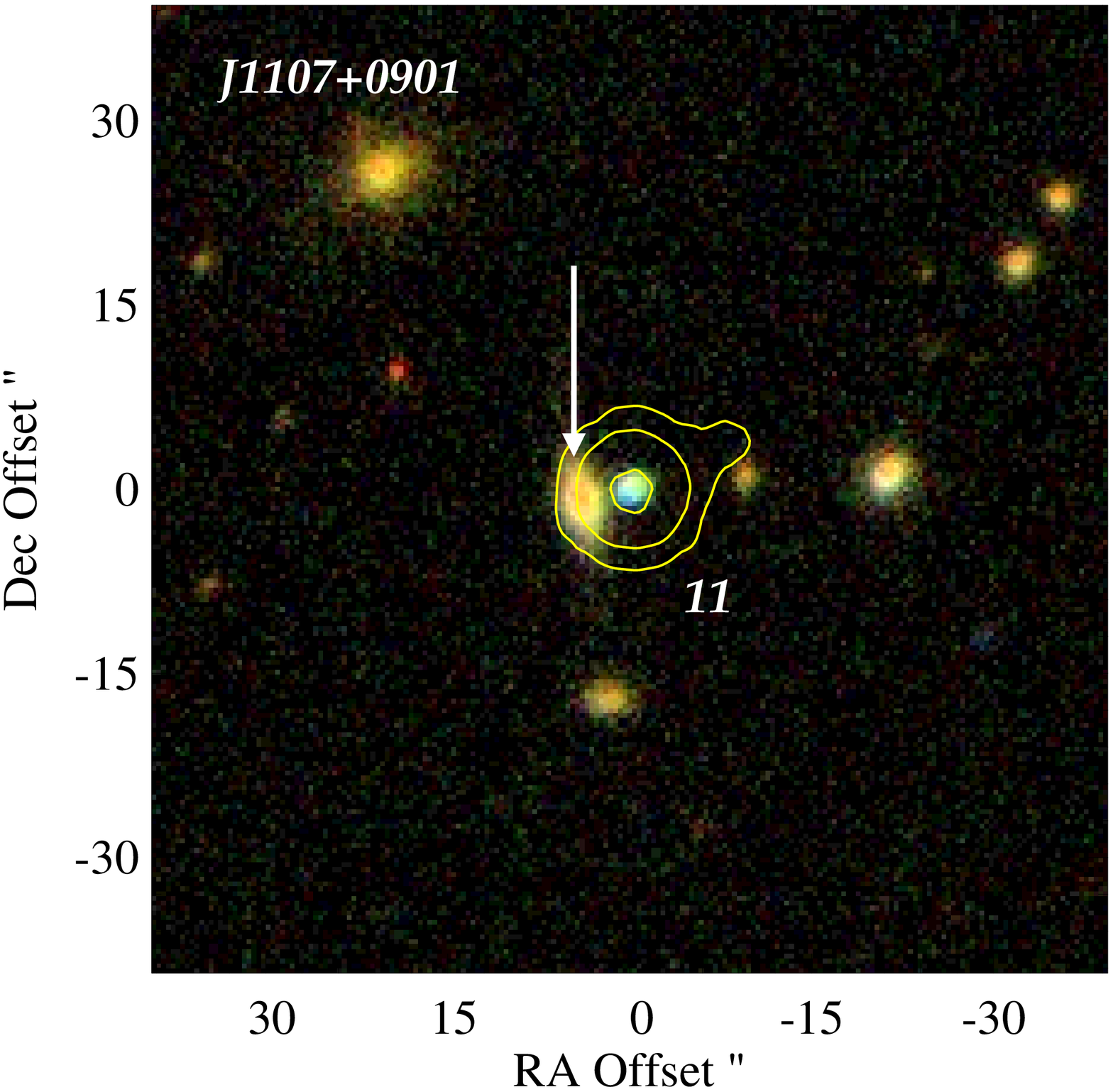} & \includegraphics[scale=0.275,angle=-0]{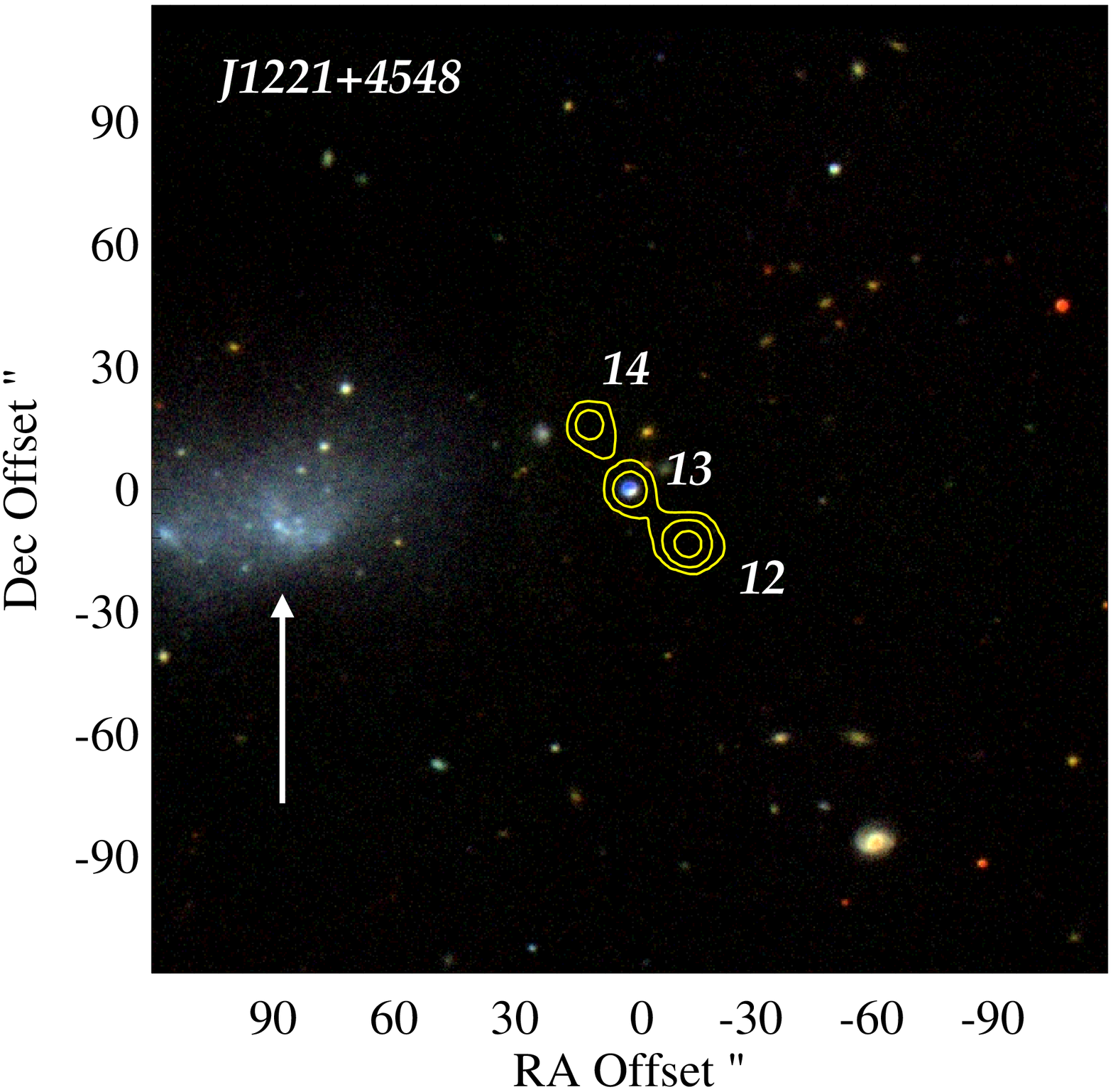} \\
\includegraphics[scale=0.275,angle=-0]{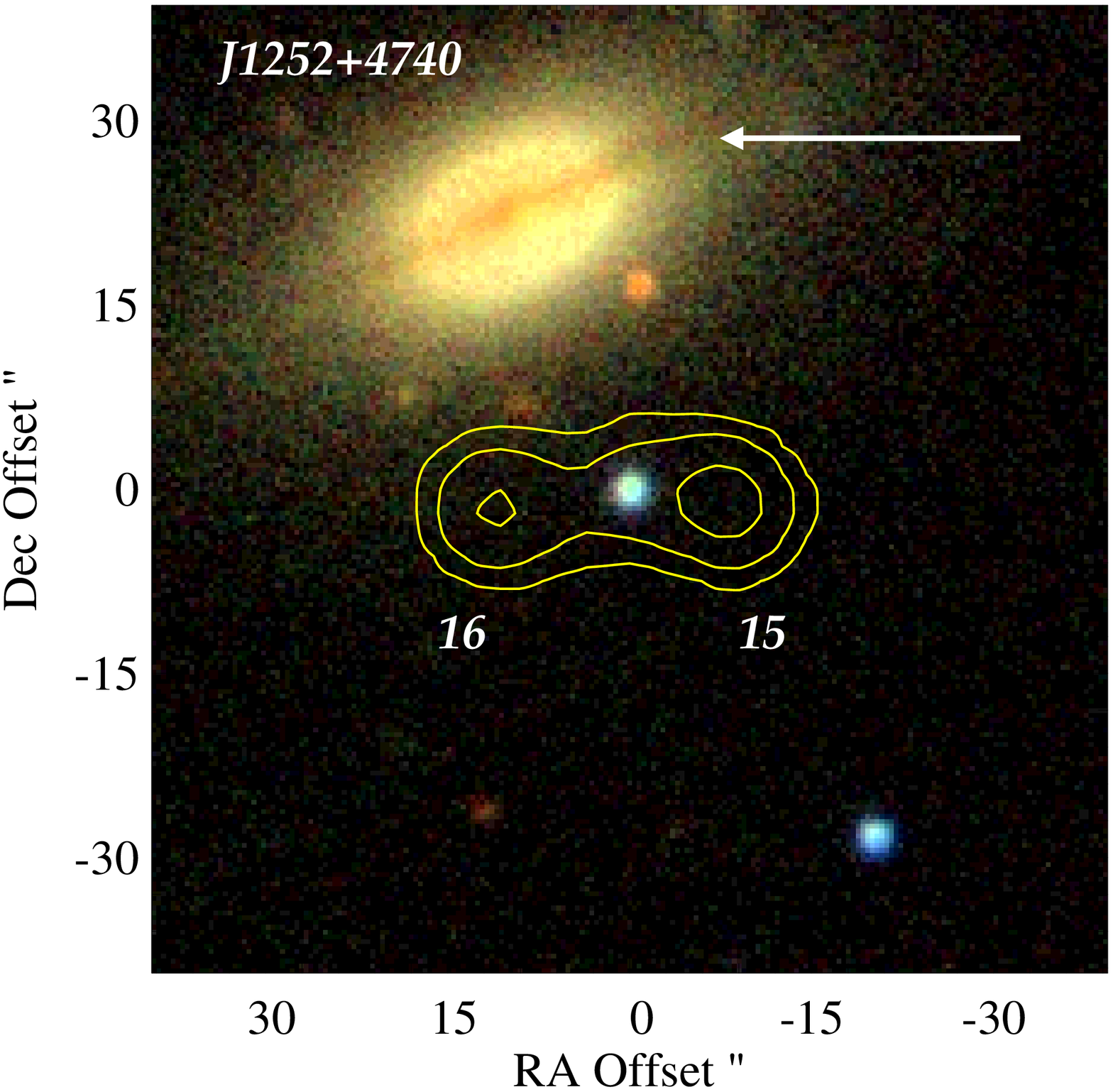} & \includegraphics[scale=0.275,angle=-0]{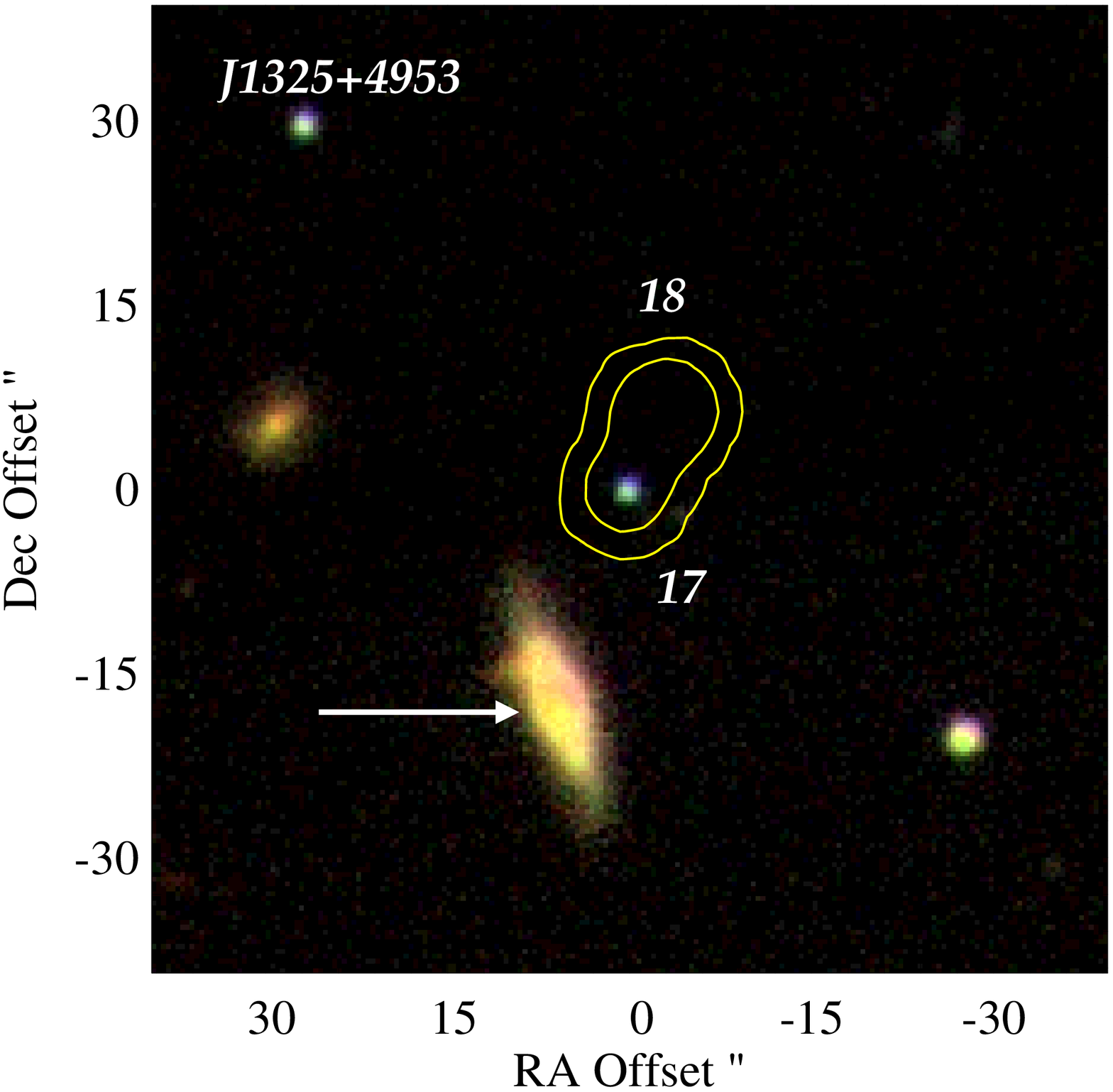} &  \includegraphics[scale=0.275,angle=-0]{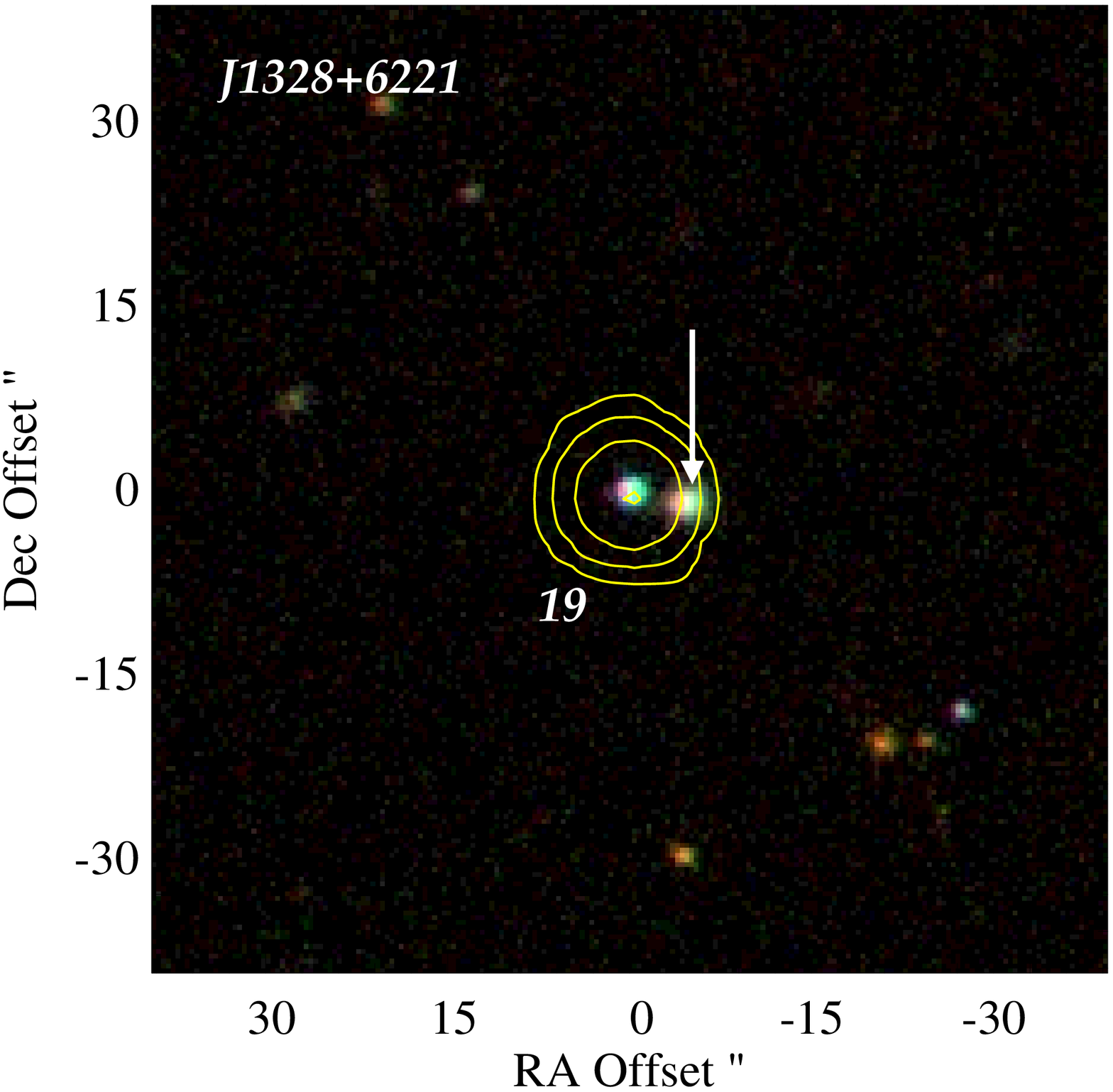} \\
\end{tabular}
{\caption{SDSS composite image of i-, r-, and g-band images for our sample of 15 galaxy-quasar pairs. The quasar is at the center of the image and its name is printed at the top left side of each plot. The yellow contours shows the 1.4~GHz VLA B-configuration continuum emission from the FIRST survey at levels 1, 5, 25, and 125~mJy/beam. The foreground galaxies are marked with a white arrow and each of the sightlines is numbered. Due to the large extent of the background quasars in 1.4~GHz some of our systems covered multiple sightlines. Details of our sample are presented in table~\ref{tbl-sample}. }\label{fig-sample}}
\end{figure*}

 \begin{figure*}
 \figurenum{1}
\begin{tabular}{l l l }
\includegraphics[scale=0.275,angle=-0]{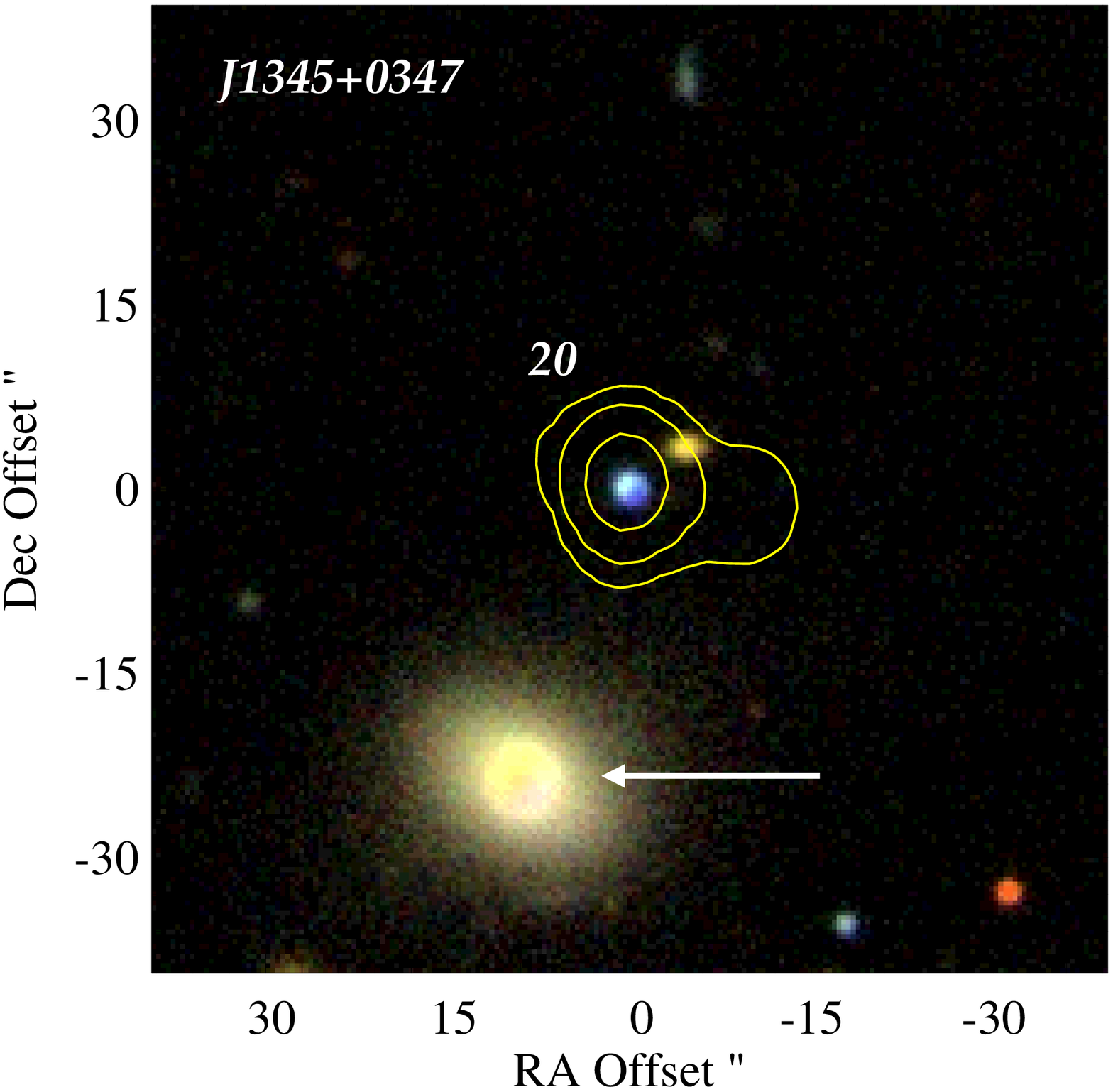} & \includegraphics[scale=0.275,angle=-0]{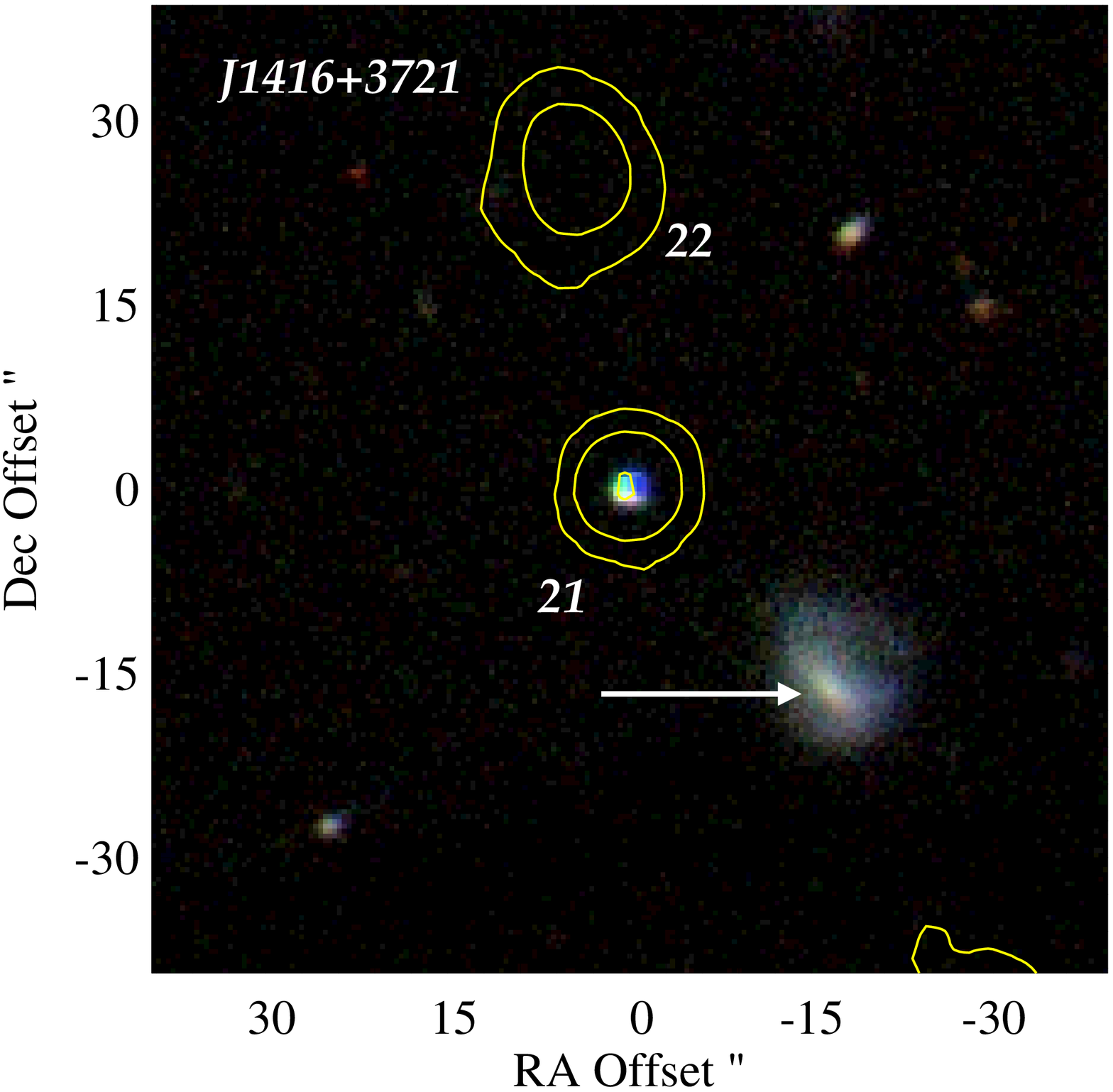} &  \includegraphics[scale=0.275,angle=-0]{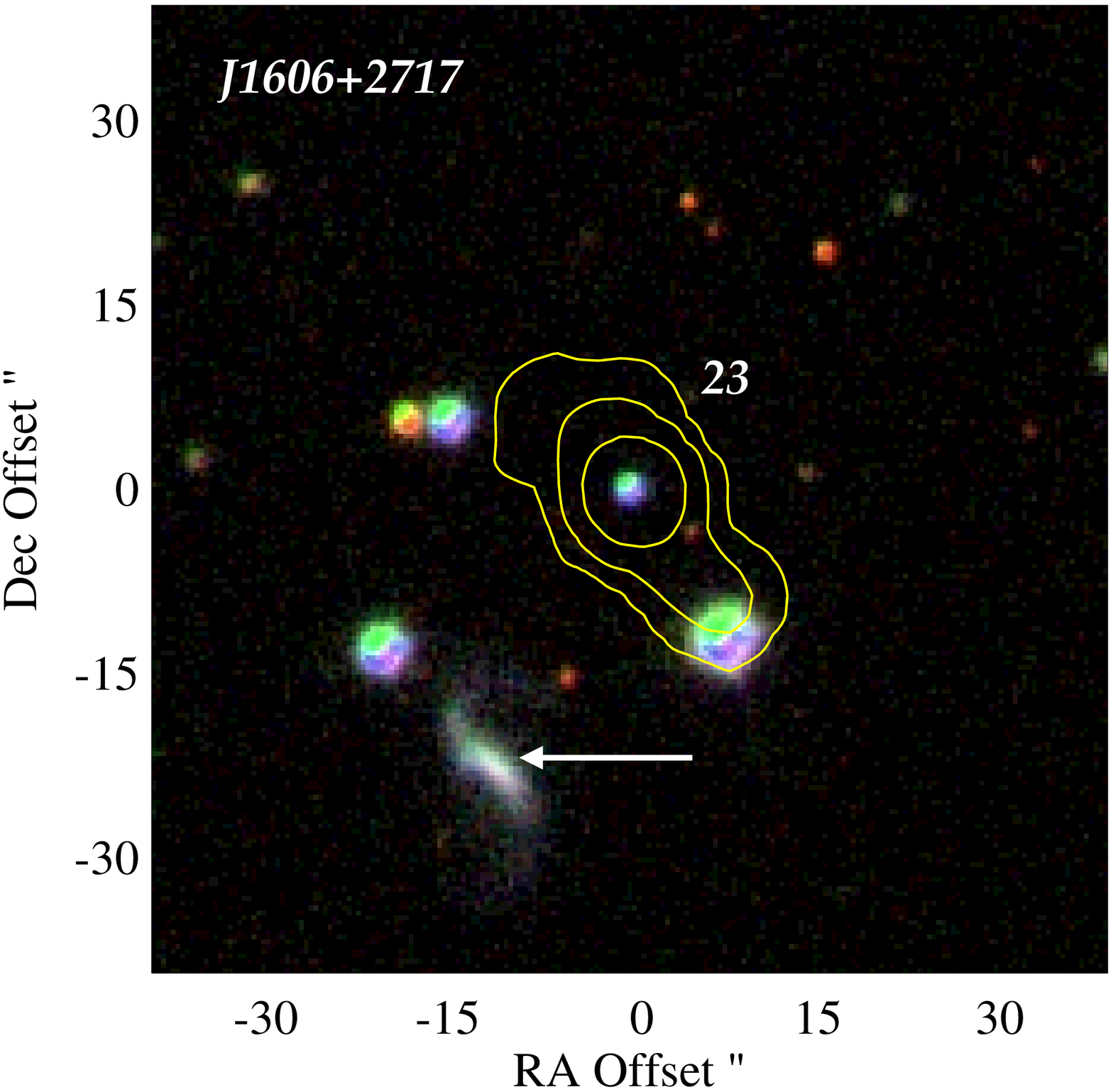} \\
\end{tabular}
{\caption{Continued.} }
\end{figure*}

We also wish to investigate the \HI distribution in terms of galaxy luminosity, mass, and color. This goal is motivated by the fact that the intensity of accretion and outflow processes are expected to vary from galaxy to galaxy. For instance, blue star-forming galaxies are expected to have a significant reserve of gas compared to red galaxies. Similarly, cold-mode accretion is predicted to be predominant in galaxies of baryonic mass $< \rm 10^{10.3}~\rm M_{\odot}$. Secondly, we intend to study the properties of \HI\ clouds, such as their size, filling factor, and temperature, in various regions of an individual galaxy. 
We begin by presenting our sample and elaborating on the selection procedure, followed by our  \HI\ 21~cm observations in section~2. 
In section~3 
we present our results and discuss their implications in section~4. 
Finally, we summarize our findings and conclusions in section~5. 
Throughout the paper we use $H_0 =70~{\rm km~s}^{-1}~{\rm Mpc}^{-1}$, $\Omega_m = 0.3$, and $\Omega_{\Lambda} = 0.7$.

\section{OBSERVATIONS  \label{sec:observations}}

\subsection{Sample Selection \label{sec:sample}}

Our sample was selected in a two-stage procedure. First, a set of background QSO sightlines close to foreground galaxies were identified by cross-correlating galaxy positions with quasar positions from objects observed by the {\it Sloan Digital Sky Survey} (SDSS). The SDSS has redshift information for some of the galaxies, while others have photometric redshift determinations. Hence to identify quasar-galaxy pairs at various impact parameters from the entire SDSS catalog, we used one of the four criteria listed below.
\begin{itemize}	
\item[1.] For SDSS galaxies with spectroscopic redshifts between $0.005 < z_{\rm{gal}} \leq 0.2$, we searched for SDSS QSOs that had sightlines passing within $20 \arcsec$ of a galaxy, or, within $1.2\times r_P (r)$, where $r_P (r)$ is the $r-$band Petrosian radius of a galaxy in arcsec. This latter selection was useful for finding low redshift galaxies where the QSO-galaxy impact parameter was $> 20 \arcsec$ but where the sightline still passed close to the optical extent of a galaxy. Similarly, we also searched for QSO sightlines from galaxies separated by $< 30 \arcsec$ but where the galaxy had $r_P (r) > 8.0$, in order to find very low redshift galaxies (with, therefore, large $r_P (r)$ on the sky) and where larger angular separations still translated to relatively small physical impact parameters.  In order to remove galaxies whose absorption features might be confused with the AGN activity of the QSO, we excluded galaxies whose redshifts were within $\Delta z =0.1$ of the QSO's redshift.
\item[2.]  We searched for SDSS QSO sightlines that passed within $30 \arcsec$ of SDSS galaxies that had {\it no} measured redshifts, with the intention of obtaining redshift information using available telescopes. We discuss these data in Section~\ref{sec:optical}.
\item[3.] We also searched for SDSS QSOs within $1.5~D_{25}$ of all galaxies in the {\it Third Reference Catalogue of Bright Galaxies} (RC3) \citep{devau91}, where D$_{25}$ is the major isophotal diameter of a galaxy measured to a surface brightness level of $\mu_B = 25.0$ square mags. 
\item[4.] For probing the smallest impact parameters, we chose foreground galaxies that were identified by detecting emission lines superimposed on the background QSO spectrum \citep{quash08,bor10}.
\end{itemize} 

Based on the above criteria, we selected about a thousand QSO-galaxy pairs. In the second stage of selecting suitable QSO-galaxy pairs, we extracted the radio flux densities associated with the optical QSOs from the {\it Faint Images of the Radio Sky at Twenty-Centimeters} (FIRST) Survey \citep{becker95}. We chose a limiting 1.4~GHz flux density value of 25~mJy for the background quasar, corresponding to 3~$\sigma$ optical depth of 0.35 for 2 hours of on-source integration for our observational settings described in the next section. Based on the observable redshift range and excluding pairs that were being observed by \citep{gupta10}  with the GMRT, this flux density cut resulted in a sample of  18 galaxy-quasar pairs.

{
\begin{sidewaystable*} 
\vspace{-7cm}
    \centering
\caption{A Detailed Description of Our Sample with Information on the Background Optical QSOs, Their Radio Counterparts, and the Foreground Galaxies.  \label{tbl-sample}}
       \begin{tabular}{ccccccccccccccc}
\hline
\hline
SL &QSO$_{opt}$ & QSO$_{radio}$ & $S_{\rm 1.4~GHz}$ &  Galaxy & $z_{gal}$ & Telescope& m$_r$ & L$_r$ &M$_{stellar}$ & Color  & $\rho$  \\

&&&(mJy)&  & & & &  (L$_*$)  & (Log[M$_{\odot}$]) &  & (kpc) &  \\
\hline

1&J010644.15-103410.5 & 010644.124-103410.55 & 265.90 & J010643.94-103419.3 & 
0.1460&APO & 18.68&0.571 & 10.0 & Blue & 23.4\\
2&J074841.77+173456.6 & 074841.773+173456.82 & 41.97 &  J074842.58+173450.6 & 
0.0528&SDSS & 16.30&0.595 & 10.6 & Red & 13.5\\
3&J074841.77+173456.6 & 074841.786+173512.20 & 21.94 &   J074842.58+173450.6 & 
0.0528&SDSS & 16.30&0.595 & 10.6 & Red & 25.1\\
4&J074841.77+173456.6 & 074842.084+173443.37 & 16.08 &    J074842.58+173450.6 & 
0.0528&SDSS & 16.30&0.595 & 10.6 & Red  &   10.4\\
5&J082153.82+503120.4 & 082153.833+503120.57 & 53.94 &   J082153.76+503125.7 & 
0.1835\tablenotemark{a}&APO & 19.60&0.418 & 10.0 & Blue & 16.0\\
6&J084914.27+275729.7 & 084914.282+275729.90 & 53.94 &   J084912.42+275740.4 & 
0.1948&APO & 18.67&1.193 & 10.5 & Blue & 86.7\\
7&J091011.01+463617.8 & 091011.016+463617.87 & 163.10 &  J091010.55+463633.3 & 
0.0998&APO & 16.94&1.286 & 10.6 & Blue & 29.8\\
8&J102258.41+123429.7 & 102258.415+123426.26 & 93.98 &  J102257.92+123439.1 & 
0.1253&APO & 18.21&0.743 & 10.7 & Red & 33.1\\
9&J102258.41+123429.7 & 102258.552+123439.94 & 24.67 &  J102257.92+123439.1 & 
0.1253&APO & 18.21&0.743 & 10.7 & Red & 20.9\\
10& J104257.58+074850.5 & 104257.598+074850.60 & 381.57 & \galaxyname\tablenotemark{c} & 
0.0332 & SDSS &18.05 &0.048 & 9.1  & Blue & 1.7 \\
11&J110736.61+090114.8 & 110736.607+090114.72 & 36.97 &   J110736.88+090113.6 & 
0.1050&APO & 17.82&0.683 & 10.7 & Red & 8.1\\
12&J122106.87+454852.1\tablenotemark{c} & 122105.480+454838.80 &  53.69 & J122115.22+454843.2 & 
0.0015&SDSS & 14.72&0.003 & 7.4 & Blue & 3.3\\
13&J122106.87+454852.1\tablenotemark{c} & 122106.854+454852.16 & 21.46 &J122115.22+454843.2 & 
0.0015&SDSS & 14.72&0.003 & 7.4 & Blue & 2.8\\
14&J122106.87+454852.1\tablenotemark{c} & 122107.811+454908.02 & 12.82  & J122115.22+454843.2 & 
0.0015&SDSS & 14.72&0.003 & 7.4 & Blue & 2.6\\
15&J125248.28+474043.7 & 125247.588+474042.81 & 78.72 & J125249.20+474105.9 & 
0.0321&APO & 13.92&1.861 & 11.1 & Red & 18.1\\
16&J125248.28+474043.7 & 125249.326+474042.19 & 45.06  & J125249.20+474105.9 & 
0.0321&APO & 13.92&1.861 & 11.1 & Red & 15.2\\
17&J132534.55+495341.9 & 132534.240+495348.47 & 26.63  & J132535.13+495324.4 & 
0.0478&SDSS & 16.27&0.497 & 10.5 & Red & 23.9\\
18&J132534.55+495341.9 & 132534.565+495342.26 & 15.87  & J132535.13+495324.4 & 
0.0478&SDSS & 16.27&0.497 & 10.5 & Red & 17.5\\
19&J132840.57+622137.0 & 132840.599+622136.65 & 144.10  &J132839.89+622136.0 & 
0.0423&APO & 18.76&0.038 & 9.0 & Blue & 4.2\\
20&J134528.74+034719.6 & 134528.765+034720.09 & 70.21 &  J134525.30+034823.8 & 
0.0325&SDSS & 14.22&1.431 & 10.9 & Red & 53.3\\
21&J141630.67+372136.8 & 141631.039+372203.01 & 30.71 &  J141629.25+372120.4 & 
0.0341&APO & 17.19&0.103 & 9.3 & Blue & 32.4\\
22&J141630.67+372136.8 & 141630.672+372137.09 & 30.53 & J141629.25+372120.4 & 
0.0341&APO & 17.19&0.103 & 9.3 & Blue & 16.2\\
23&J160658.30+271705.5 & 160658.315+271705.86 & 141.01   &J160659.13+271642.6 & 
0.0462&APO & 17.82&0.111 & 9.3 & Blue & 23.3\\
- &J082153.82+503120.4 & 082153.833+503120.57 & 53.94 &  J082152.85+503128.4 & 
0.1694\tablenotemark{a}&APO & 18.53&1.066 & 10.8 & Red & 35.3\\
- &J084914.27+275729.7 & 084914.282+275729.90 & 53.95 &  J084913.80+275735.9 & 
0.2354&APO & 18.69&1.886 & 10.9 & Blue & 32.8\\
- &J115839.90+625428.0 & 115836.882+625415.06 & 401.20 & J115840.67+625423.7 & 
0.2596&SDSS & 18.49&2.887 & 11.1 & Blue & 109.7\\
- &J115839.90+625428.0 & 115837.852+625420.28 & 180.07 &J115840.67+625423.7 & 
0.2596&SDSS & 18.49&2.887 & 11.1 & Blue & 78.6\\
- &J115839.90+625428.0 & 115842.780+625442.26 & 69.64 &J115840.67+625423.7 & 
0.2596&SDSS & 18.49&2.887 & 11.1 & Blue & 94.5\\
\hline
\multicolumn{5}{l}{\footnotesize $^a$ Gupta et al. 2010 in press}\\
\multicolumn{8}{l}{\footnotesize $^b$  Identified through emission lines on the QSO spectrum \citep[see][]{quash08,bor10} }\\ 
\multicolumn{5}{l}{\footnotesize $^c$ Also known as UGC~7408.}\\
             \end{tabular}
\end{sidewaystable*} }

We carried out observations of the 18 galaxy-pairs with the {\it Green Bank Telescope} (GBT). We encountered severe radio frequency interference (RFI) for three of the pairs, thus reducing our sample to 15 galaxy-quasar pairs. Some of the background quasars are extended radio sources with multiple peaks at the resolution of the FIRST Survey, and therefore allow us to probe different regions of the foreground galaxies. The  GBT spectrum for such a pair would have contributions from each of the peaks seen in FIRST image as the GBT beam encompasses all the peaks of the extended quasars. Although, this would cause confusion in the case of a detection, however, this is particularly advantageous for nondetections. More stringent limits can be estimated for each of the background continuum peaks for nondetections. 6 of our 15 galaxy-quasar pairs have extended quasars and these GBT nondetections essentially probe 14 slighlines. Hence, we present our data for the 23 sightlines through 15 galaxy-quasar pairs.

Details on the sightlines along with their 1.4~GHz radio flux densities are presented in Table~\ref{tbl-sample}. Radio contours from the FIRST survey overlaid on the composite SDSS {\it r}, {\it g}, and {\it i}-band images for our sample are presented in Figure~\ref{fig-sample}. The QSO is centered in each image and the foreground galaxy is marked by a white arrow. The QSO name is printed at the top left-hand side of the image whereas the sightline numbers used in Table~\ref{tbl-sample} are reproduced near the galaxy or background radio source.
 Including multiple radio peaks, our sample covers a wide range of impact parameters from 1.7 to 86.7~kpc as well as a wide variety of galaxy properties (see next section). The sightline toward the background QSO \qsoname\ allows us to probe the inner stellar disk of the foreground dwarf spiral galaxy \galaxyname\ at an impact parameter of 1.7~kpc (see \citet{bor10} for details). Besides a wide continuous range of impact parameters, we also cover a wide range of galaxy properties including luminosity, mass, and color.

\subsection{Optical Properties and Galaxy redshifts \label{sec:optical}}

For foreground galaxies without redshift information from SDSS, we conducted spectroscopic observations from the Apache Point Observatory (APO) to obtain their redshifts. The APO redshifts were measured either by fitting Gaussian profiles to emission lines or from the stellar absorption line measurements made using the Image Reduction and Analysis Facility (IRAF) ``fxcor'' routine, and cross-correlating the blue spectra with that of a radial velocity standard. The redshifts are presented in the sixth column of Table~\ref{tbl-sample} with more details in Table~\ref{tbl-redshift}. With the exception of two galaxies, the remaining APO redshifts are measured using absorption-lines.
The r-band magnitude and luminosity for the foreground galaxies are presented in the eighth and ninth columns of Table~\ref{tbl-sample}. The luminosities, in units of $L^*$, were estimated using $M^{*}_{r} = -21.2$  \citep[Table 2 of][]{blanton03} and were k-corrected for r-band absolute magnitude at $z=0.0$ . The k-corrections were derived using the k-correct algorithm by \citet{blanton08} that uses SDSS photometric u, g, r, i, and z-band data. 
Due to the unavailability of SDSS photometric data for \galaxyname\, we have used $g, r,$ and $i-$band photometric values from \citet{bor10}  to estimate the k-correction for this galaxy. We also present stellar masses for the foreground galaxies, which were estimated using Equation~(1) from \citet{mcintosh08} based on the k-corrected magnitudes.

\begin{table}
\caption{APO redshifts for Foreground Galaxies Using Emission and Absorption lines.  \label{tbl-redshift}}
\centerline {
\begin{tabular}{cccc }
\hline\hline
Foreground Galaxy & $z_{em-line}$ & $z_{abs-line}^a$ & $\delta z_{abs-line}^a$  \\
\hline
J010643.94-103419.3 &	0.1460 &      -  &   -  \\	 			
J084913.80+275735.9 &	0.2356 &  0.2354 & 0.0004 \\ 
J084912.42+275740.4 &   0.1941 &  0.1948 & 0.0008 \\ 
J091010.55+463633.3 &	0.0998 &  0.0998 & 0.0004 \\ 
J102257.92+123439.1 &	0.1254 &  0.1253 & 0.0004 \\ 
J110736.88+090113.6 &	  -        &  0.1050 & 0.0003 \\ 
J132839.89+622136.0 &	0.0420 &  0.0423 & 0.0005 \\ 
J141629.25+372120.4 &   0.0341 &    -    & -    \\	 	 	\hline
\multicolumn{2}{l}{\footnotesize $^a$ Estimated using IRAF fxcor routine.}\\
\end{tabular}
\\}
\end{table}

From the galaxy color-magnitude diagram that relates galaxy luminosity and color one can classify galaxies as part of the red sequence or the blue cloud. For our sample, we follow the empirical results from \citet[][equation A6]{vandenbosh08}, which uses the parameter $^{0.0}A$ defined as
\begin{equation}
^{0.0}A=^{0.0}(g-r)-0.10~(\rm log[M_*]-10.0) 
\end{equation}
where $^{0.0}(g-r)$ is based on {\it g} and {\it r} band magnitudes that have been k-corrected to $z=0.0$ and $M_*$ is the stellar mass of the galaxy.
The two populations are found to have an overlap in terms of the $^{0.0}A$ value intersecting at $^{0.0}A=0.675$ \citep[see Figure A1 of][]{vandenbosh08}. Hence we use this value to distinguish the two populations and we classify galaxies with $^{0.0}A<0.675$ as belonging to the blue cloud and ones with $^{0.0}A>0.675$ as red sequence.  In total, our sample consists of nine blue cloud and six  red sequence galaxies.

\subsection{ Radio Observations \label{sec:radio_21cm}}

In order to search for cold gas in the outer disks and halos of galaxies, we carried out GBT observations of our sample followed by higher spatial resolution Very Large Array (VLA) and Very Large Baseline Array (VLBA) observations. Details of the VLA and VLBA observations of the pair \qsoname -\galaxyname\ have been presented in \citet{bor10} and are not presented in this section.

\subsubsection{21 cm Spectroscopy with the GBT \label{sec:gbt_21cm}}

We conducted GBT observations of the \HI\ 21~cm and OH 18~cm transitions for 18 galaxy-quasar pairs (Table~\ref{tbl-sample}) under project IDs GBT06B-052, GBT08A-082, and GBT09C-039 in August 2006, March through July 2008, and October and November 2009 respectively. We used the dual polarization L-band system with 9-level sampling. The intermediate frequency (IF) was set to yield a channel width of  1.56~kHz (0.33~km s$^{-1}$) using 8192 channels over a total bandwidth of 12.5~MHz. Observations were made in the standard position-switching scheme by cycling through the ON-OFF sequence, then dwelling for 300~s at each position. Each 300~s scan consisted of thirty 10~s integrations to minimize data corruption by sporadic RFI. The OFF position was chosen +20\arcmin\ offset in Right Ascension from each of the sightlines. This was done to track the possible presence of a source in the OFF position. During each of the observing sessions, we used one of the three standard flux calibrators - 3C~48 (16.5~Jy), 3C~147 (22.5~Jy), and 3C~286 (15.0~Jy) - for pointing and estimating antenna gain. Local pointing corrections (LPC) were performed using the observing procedure AutoPeak and the corrections were then automatically applied to the data. This should result in a pointing accuracy of 3$^{\prime\prime}$. Similarly, flux calibrations were performed by applying antenna gain to data for each session separately (values are presented in Table~\ref{tbl-gbt_obs}). The corrections also include attenuation due to opacity of the air (around 1\%).

The data were analyzed using the NRAO software GBTIDL. Most of the sightlines were observed for almost 2 hours on-source. However,  due to the corruption of the data by RFI the noise levels achieved for each of the sightlines span a wide range. Severe RFI corrupted entire datasets for sightlines toward the background radio QSOs 082153+503120 at $z=0.1640$, 084914+275729 at z=0.2357, and 115836+625415 at z=0.2596 rendering the data unusable. Besides RFI, the GBT data also suffered from standing waves for some of the brighter sources, which limited our \HI\ emission detection capabilities, especially for \HI\ features with $\Delta~V \ge$ 300~\kms. The details of the noise properties and GBT measurements are presented in Table~\ref{tbl-gbt_obs}. No OH emission was detected in any of our sources and rms noise levels are presented in the above mentioned table.

 \begin{figure}
 \figurenum{2}
\begin{tabular}{c }
\includegraphics[trim = 10mm 0mm 4mm 0mm, clip,scale=0.7,angle=-0]{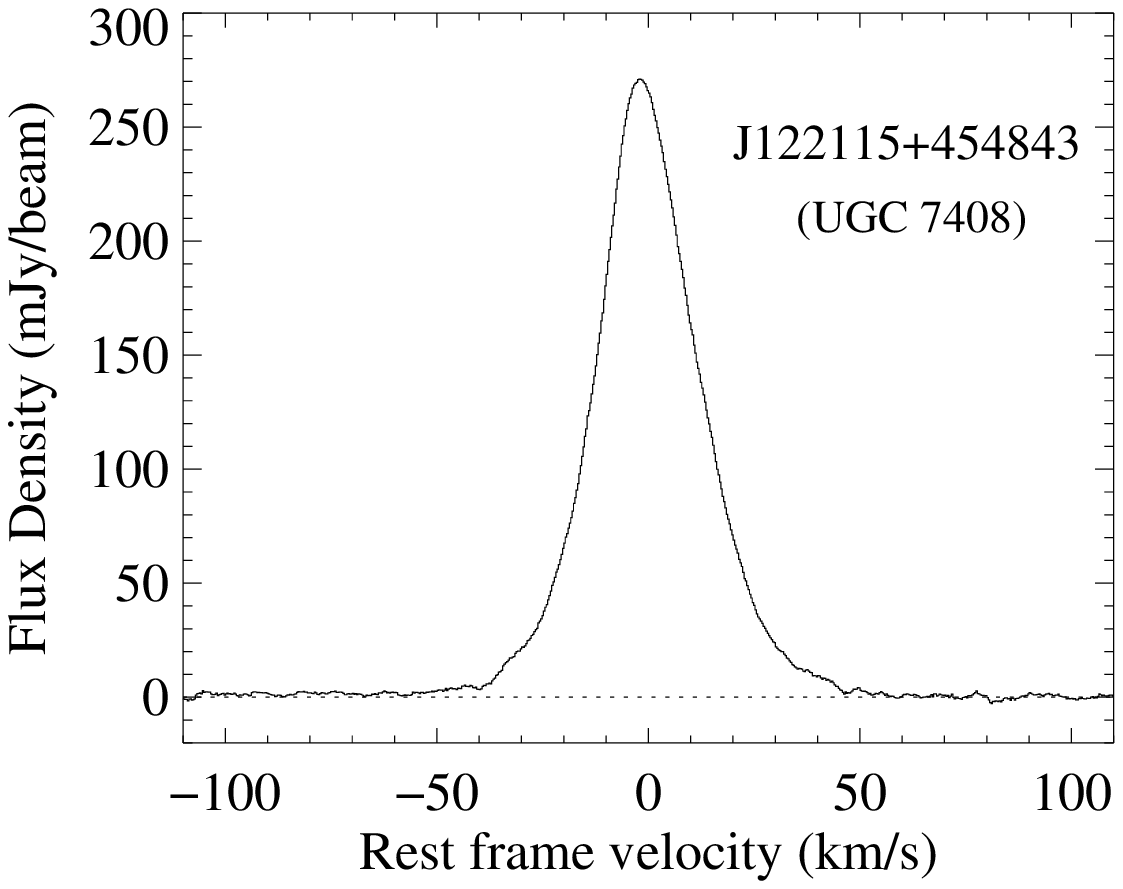} \\
 \includegraphics[trim = 10mm 0mm 5mm 0mm, clip,scale=0.7,angle=-0]{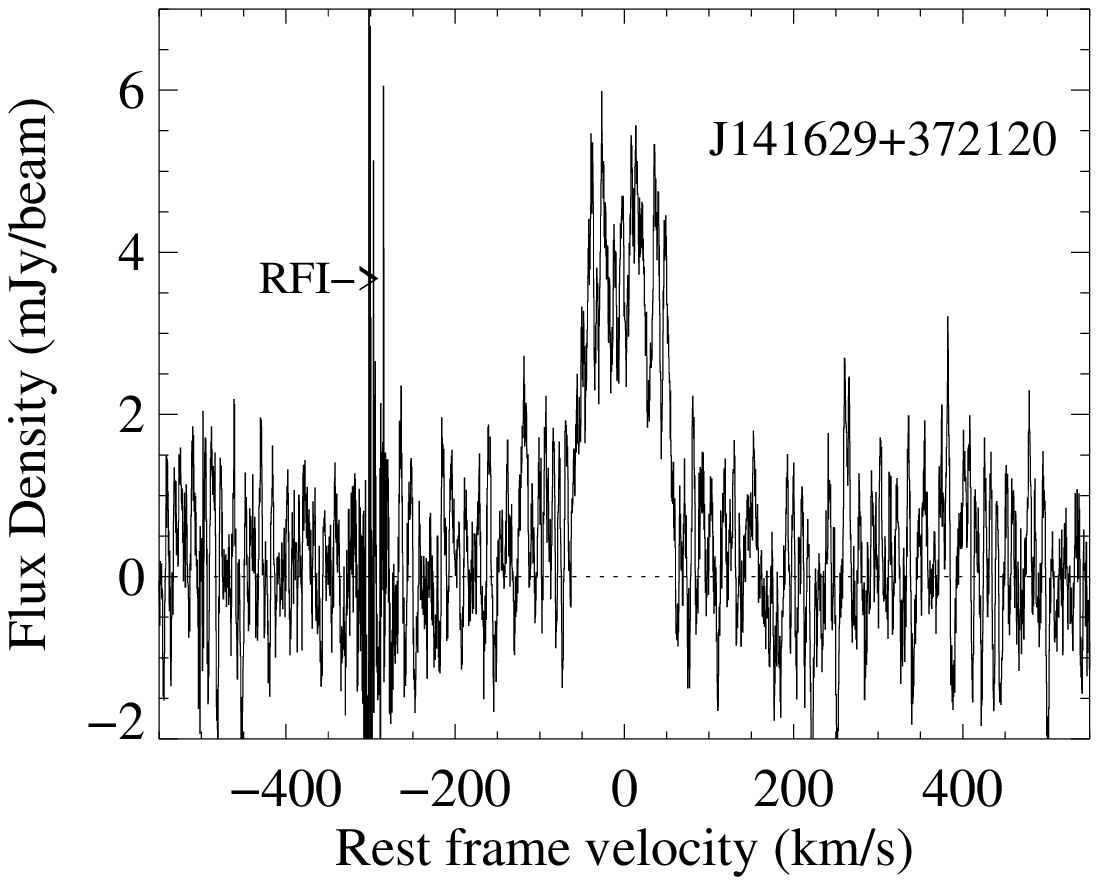} \\
 \includegraphics[trim = 10mm 0mm 5mm 0mm, clip,scale=0.7,angle=-0]{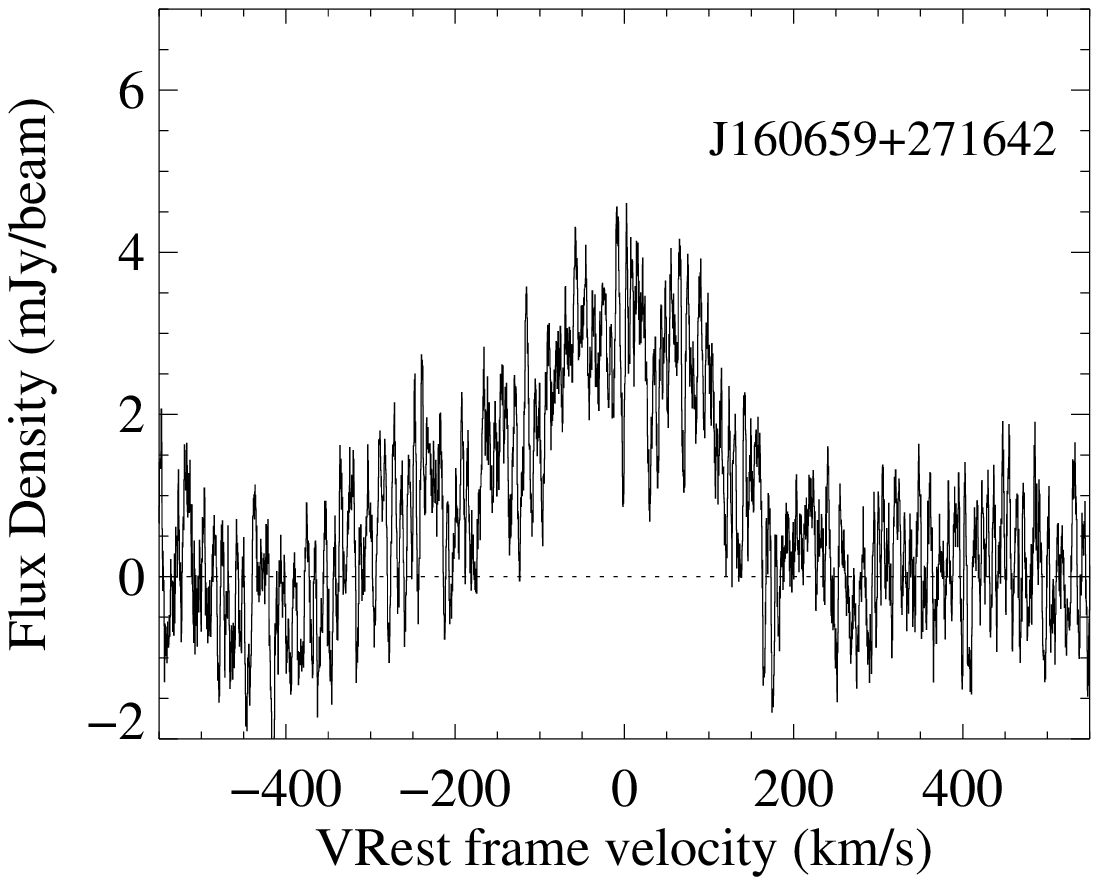}  \\
\end{tabular}
{\caption{\HI\ spectra for foreground galaxies UGC~7408 (top), J141629+372120 (center), and J160659+271642 (bottom). The spectra are presented at the rest frame of the optical galaxies to show the match between the optical and \HI\ velocity distribution. All three spectra have been smoothed to 3~\kms (i.e. 10 pixels). The \HI\ mass associated with each of these sources are presented in Table~\ref{tbl-gbt_obs}.}}
\label{hi_emm} 
\end{figure}

\subsubsection{Very Large Array Observations of J122115+454838  \label{sec:obs_vla}}

In order to understand the effect of the large GBT beam size on the survey, we imaged the \HI\ distribution in the foreground galaxy J122115+454838 (UGC~7408) with the VLA under program AY190 on 28 June and 16 July 2008. The observations were carried out in the VLA D-configuration for 3.5 hours of on-source integration. We used dual polarization using 2 intermediate frequency (IF) correlator modes to achieve a total bandwidth of 1.5625~MHz (330~\kms) with a 12.2~kHz resolution after Hanning smoothing. This correlator setup allowed us to cover the entire \HI line-width with a velocity resolution of 2.6 km~s$^{-1}$. The data were calibrated following the standard VLA calibration and imaging procedures using NRAO's Astronomical Image Processing Software (AIPS). The data were imaged using both natural and uniform weighting schemes. The natural weighting was applied to enhance the fainter emission features whereas uniform weighting was applied to achieve a higher spatial resolution for the purpose of extracting a spectrum with minimum contamination from the \HI\ emitting region. The synthesized beam produced using natural weighting was 67.51$^{\prime\prime}$ $\times$ 45.31$^{\prime\prime}$ (Position angle= $-86.75^\circ$) and uniform weighting was 52.17$^{\prime\prime}$ $\times$ 38.11$^{\prime\prime}$ (Position angle= $-80.56^\circ$). The absolute uncertainty in the resulting flux density scaling is about 10\%, and this is the formal uncertainty we quote for all physical parameters derived from the flux density. The average rms noise achieved, after combining the two polarizations, is 1.07~mJy/beam/channel.

\begin{table*}
\caption{GBT 21~cm Observations and Measurements of \HI\ Optical Depth and Mass Associated With the Foreground Galaxies.\label{tbl-gbt_obs}}
\centerline {
\begin{tabular}{ccccccccccccc }
\hline\hline
SL & QSO$_{radio}$ & S$_{1.4 GHz}$&Gain & $\sigma_{HI}$ &$\tau^a$ & $\sigma_{OH}$& Foreground Galaxy & Distance  &  HI Mass\\
\hline
1&010644.124-103410.55 & 265.90 & 1.68 &3.65&  $<$ 0.02 &2.02& J010643.94-103419.3 & 692.4&  -  \\
2&074841.773+173456.82 & 41.97 & 1.72 &2.45&  $<$ 0.10 &1.63&J074842.58+173450.6 & 235.2&  -  \\
3&074841.786+173512.20 & 21.94 & 1.72 &2.45&  $<$ 0.19 &1.63&J074842.58+173450.6  & 235.2&  -  \\
4&074842.084+173443.37 & 16.08 & 1.72 &2.45&  $<$ 0.26 &1.63&J074842.58+173450.6  & 235.2&  -  \\
5&082153.833+503120.57 & 53.94 & 1.70 &4.96&  $<$ 0.16 &2.41&J082153.76+503125.7  & 890.5&  -  \\
6&084914.282+275729.90 & 53.94 & 1.72 &6.36&  $<$ 0.20 &1.98&J084912.42+275740.4  & 951.7&  -  \\
7&091011.016+463617.87 & 163.10 & 1.71 &1.31&  $<$ 0.01 &RFI&J091010.55+463633.3  & 459.3&  -  \\
8&102258.415+123426.26 & 93.98 & 1.70 &1.86&  $<$ 0.03 &1.88&J102257.92+123439.1 & 586.4&  -  \\
9&102258.552+123439.94 & 24.67 & 1.70 &1.86&  $<$ 0.13 &1.88&J102257.92+123439.1 & 586.4&  -  \\
10&104257.598+074850.60 & 381.57 & 1.65& 4.2  & 0.04962  & RFI &  \galaxyname\tablenotemark{b} & 145.9 & -  \\
11&110736.607+090114.72 & 36.97 & 1.82 &3.81&  $<$ 0.18 &RFI&J110736.88+090113.6 & 484.9&  -  \\
12&122105.480+454838.80 & 53.69 & 1.74 &3.49&  $<$ 0.11 &1.90&J122115.22+454843.2\tablenotemark{c}  & 8.4&1.4\\
13&122106.854+454852.16 & 21.46 & 1.74 &3.49&  $<$ 0.28 &1.90&J122115.22+454843.2\tablenotemark{c}  & 8.4&1.4\\
14&122107.811+454908.02 & 12.82 & 1.74 &3.49&  $<$ 0.47 &1.90& J122115.22+454843.2\tablenotemark{c} & 8.4&1.4\\
15&125247.588+474042.81 & 78.72 & 1.86 &2.64&  $<$ 0.06 &RFI&J125249.20+474105.9 & 140.9&  -  \\
16&125249.326+474042.19 & 45.06 & 1.86 &2.64&  $<$ 0.10 &RFI&J125249.20+474105.9 & 140.9&  -  \\
17&132534.240+495348.47 & 26.63 & 1.64 &5.90&  $<$ 0.38 &RFI&J132535.13+495324.4 & 212.2&  -  \\
18&132534.565+495342.26 & 15.87 & 1.64 &5.90&  $<$ 0.64 &RFI&J132535.13+495324.4 & 212.2&  -  \\
19&132840.599+622136.65 & 144.10 & 1.78 &3.02&  $<$ 0.04 &2.13&J132839.89+622136.0 & 187.0&  -  \\
20&134528.765+034720.09 & 70.21 & 1.91 &3.12&  $<$ 0.08 &RFI&J134525.30+034823.8 & 142.7&  -  \\
21&141631.039+372203.01 & 30.71 & 1.87 &2.75&  $<$ 0.16 &RFI&J141629.25+372120.4 & 149.9&23.2\\
22&141630.672+372137.09 & 30.53 & 1.87 &2.75&  $<$ 0.16 &RFI&J141629.25+372120.4 & 149.9&23.2\\
23&160658.315+271705.86 & 141.01 & 1.76 &2.34&  $<$ 0.03 &RFI&J160659.13+271642.6 & 204.8&81.3\\
-&082153.833+503120.57 & 53.94 &   RFI    &  RFI  &  RFI &1.88&J082152.85+503128.4 & 815.1&-\\
-&084914.282+275729.90 & 53.95 &   RFI    &  RFI  &  RFI &2.40&J084913.80+275735.9 & 1177.1&-\\
-&115836.882+625415.06 & 401.20 &   RFI    &  RFI  &  RFI &2.20&J115840.67+625423.7 & 1315.4&-\\
-&115837.852+625420.28 & 180.07 &   RFI    &  RFI  &  RFI &2.20&J115840.67+625423.7 & 1315.4&-\\
-&115842.780+625442.26 & 69.64 &   RFI    &  RFI  &  RFI &2.20&J115840.67+625423.7 & 1315.4&-\\
\hline
\multicolumn{9}{l}{\footnotesize $^a$ Integrated optical depth binned over 1 \kms bins i.e 3 pixels. The limiting values show 3~$\sigma$ l optical depth. }\\
\multicolumn{9}{l}{\footnotesize $^b$ Identified through emission lines on the QSO spectrum \citep[see][]{quash08,bor10} }\\
\multicolumn{8}{l}{\footnotesize $^c$ Also known as UGC~7408}\\
\end{tabular}}
\end{table*}

\section{Results \label{sec:results}}

We detected \HI\ emission from three foreground galaxies: J122115.22+454843.2 (UGC~7408), J141629.25+372120.4, and  J160659.13+271642.6, thus suggesting the presence of gas-rich galaxies in our sample. Figure~\ref{hi_emm} shows the smoothed GBT \HI\ spectra for these galaxies. The spectra are presented in terms of the optical velocity of the galaxies, which show very good correspondence with the \HI distribution.
The \HI\ masses presented in Table~\ref{tbl-gbt_obs} were estimated using the relationship M(HI) = 2.36$\times10^5D^2(S\Delta V)~M_\odot$, where $D$ is the luminosity distance to the galaxy in megaparsec (Mpc) and the $S\Delta V$ is the velocity integrated \HI\ flux density in Jy~\kms\ . All the three galaxies are ``blue" (as defined in Section~\ref{sec:optical}) although they vary in luminosity and stellar mass.  While J122115.22+454843.2 (UGC7408) is a gas rich dwarf galaxy of luminosity $0.003~L_*$ and \HI\ mass of  1.4$~\times~$10$^8~\rm M_{\odot}$, J141629.25+372120.4 and J160659.13+271642.6 are $\sim 0.1L_*$ galaxies and have \HI\ masses of  2.3$~\times~$10$^9~\rm M_{\odot}$ and 8.1$~\times~$10$^9~\rm M_{\odot}$, respectively.

\begin{figure*}
 \figurenum{3}
\hspace{-0cm}
\includegraphics[trim = 55mm 100mm 55mm 105mm, clip,angle=0,scale=1]{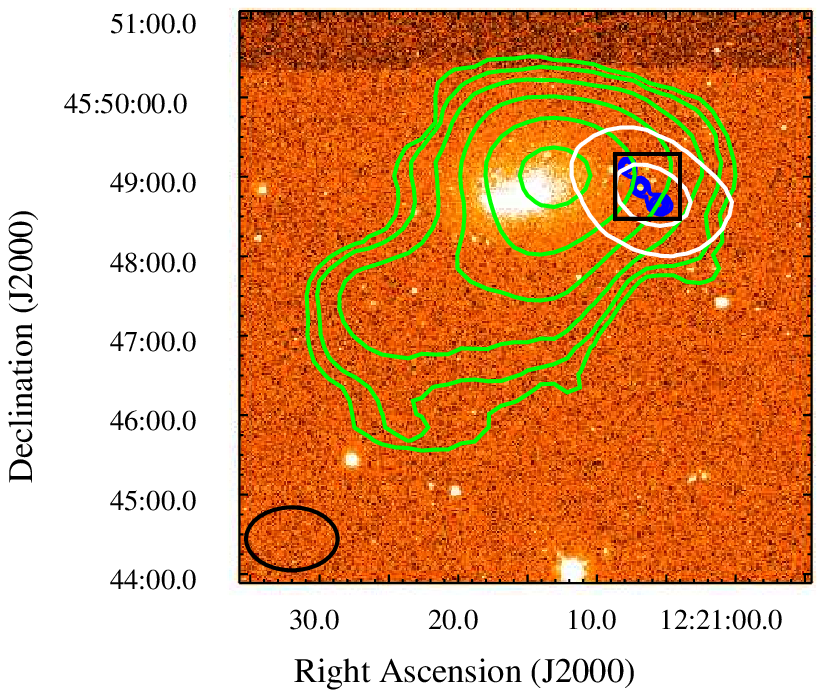} 
\hspace{-1.6cm}
\includegraphics[trim = 55mm 100mm 55mm 105mm, clip,angle=0,scale=1]{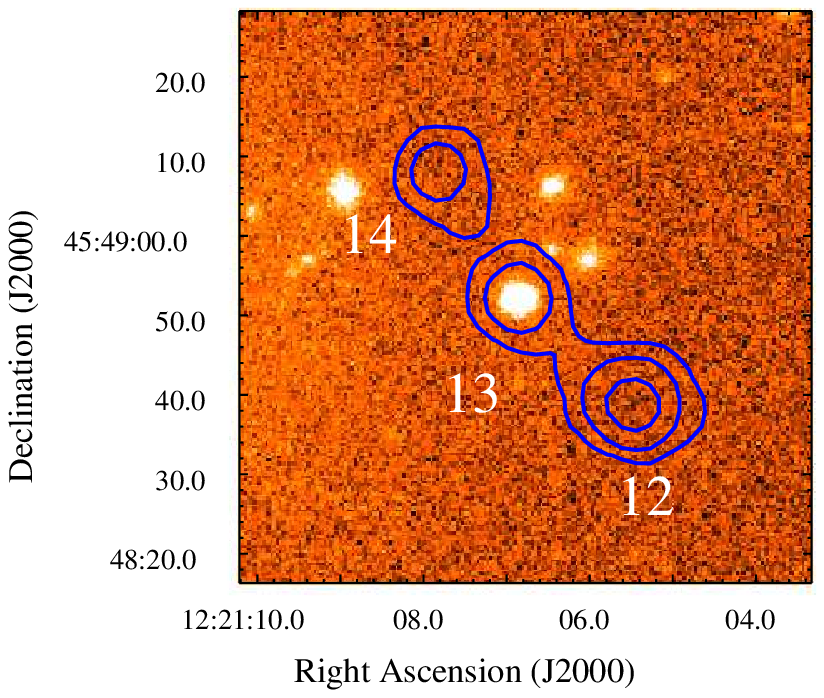}
\caption{ LEFT: VLA D-configuration \HI\ 21~cm column density contour image of UGC~7408 overlaid as green contours on an SDSS-r band image. The contours levels are at 1.8, 3.6, 7.2, 14.4, 28.8, \& 57.6~$\times $ 10$^{19} \times$ cm$^{-2}$ respectively. The beam size associated with the \HI\ data obtained using natural weighting is shown at the bottom left corner (67$^{\prime\prime}$.51 $\times$ 45$^{\prime\prime}$.31). The blue contours represent the VLA FIRST survey image of the background quasar J122106.87+454852.1 at 1, 5, and 25 mJy/beam and the white contours represent the VLA D-configuration 21~cm continuum image obtained using uniform weighting at 5 and 50 mJy/beam (Position angle= $-86.75^\circ$). The higher resolution FIRST survey images shows three peaks from the background source, which are blended together into a single source in the D-configuration image. The black square represents the region corresponding to the red spectrum in  Figure~\ref{VLA_HI_spec}, which is roughly the size of the beam for the uniformly weighted data. RIGHT: A magnified version of the left hand panel showing the background quasar as imaged by the FIRST survey. The brightest background source at 1.4~GHz is the southern component of the quasar.  The three peaks of the background quasar mark the sightlines 12, 13, and 14.
\label{VLA_HI_image}}
\end{figure*}

The VLA D-configuration image of \HI\ emission associated with UGC~7408 (green contours), overplotted on an SDSS r-band image, is presented in the left panel of Figure~\ref{VLA_HI_image}. 
The FIRST image (beam size = 5$^{\prime\prime}$.40 $\times$ 5$^{\prime\prime}$.40) of the background QSO J122106.87+454852.1 with its lobes is shown in blue  in (enlarged version in the right panel) and its D-configuration continuum image (beam size = 52$^{\prime\prime}$.17 $\times$ 38$^{\prime\prime}$.11) is shown with white contours. The sightlines (sightline (SL)\#~12, 13, and 14) to the background continuum source probes the galaxy at impact parameter of 3.2, 2.8, and 2.6~kpc. The \HI\ distribution in this galaxy is extended, extending 2-3 times beyond the optical size of the galaxy. The \HI\ is asymmetric and extends 5$^{\prime}$.6 to the southeast of the galaxy.
 {\it The Westerbork observations of neutral Hydrogen in Irregular and SPiral galaxies } \citep[WHISP Survey;][]{swaters02} also imaged this galaxy. The subsequent analysis by \citet{swaters09} found that its \HI\ distribution is clumpy and lacking ordered rotation.

\begin{figure*}
 \figurenum{4}
\hspace{-.3cm}
\includegraphics[trim = 0mm 0mm 0mm 0mm, clip,angle=0,scale=.45]{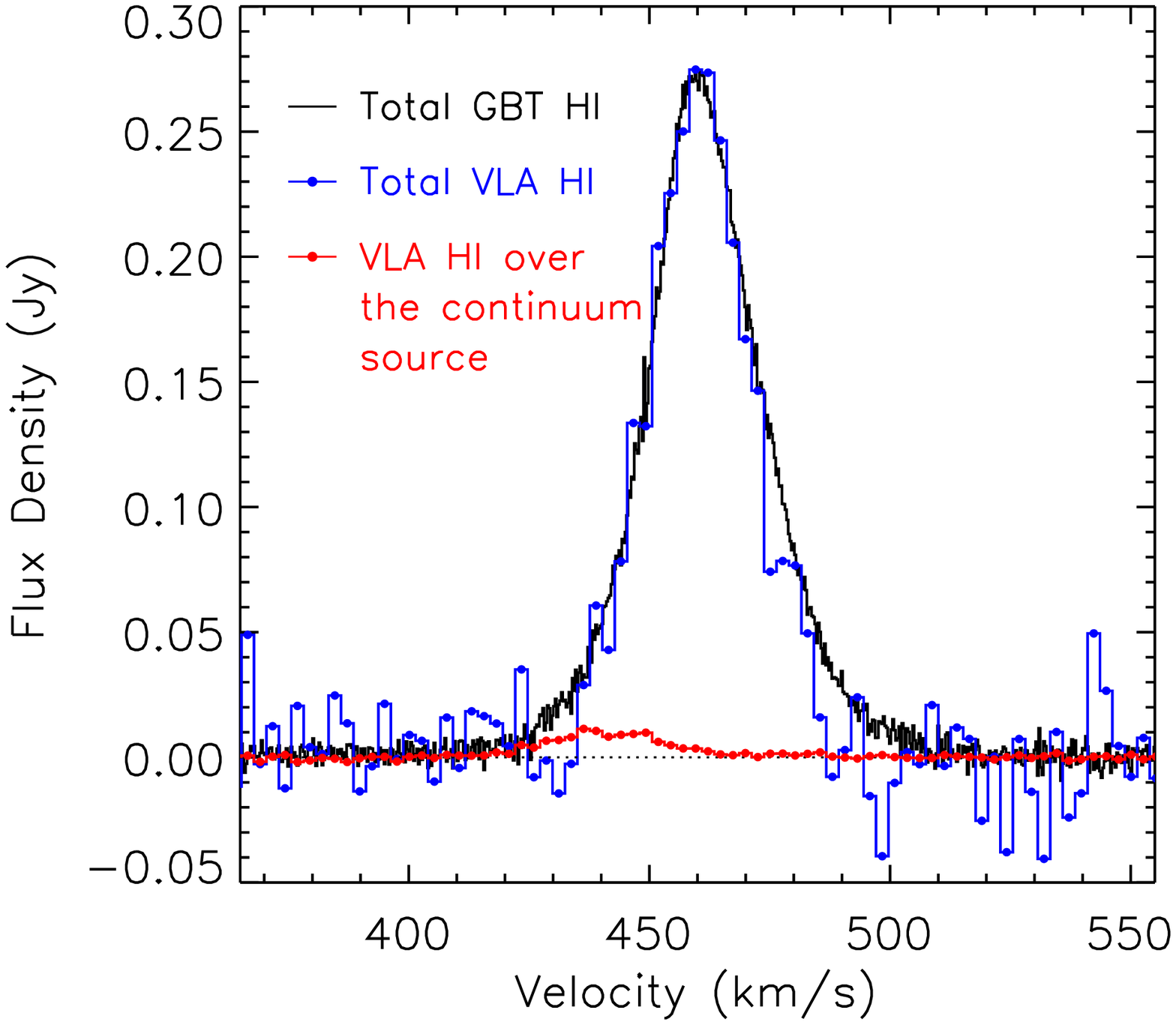} 
\hspace{-1.3cm}
\includegraphics[trim = 0mm 0mm 0mm  0mm, clip,angle=0,scale=.45]{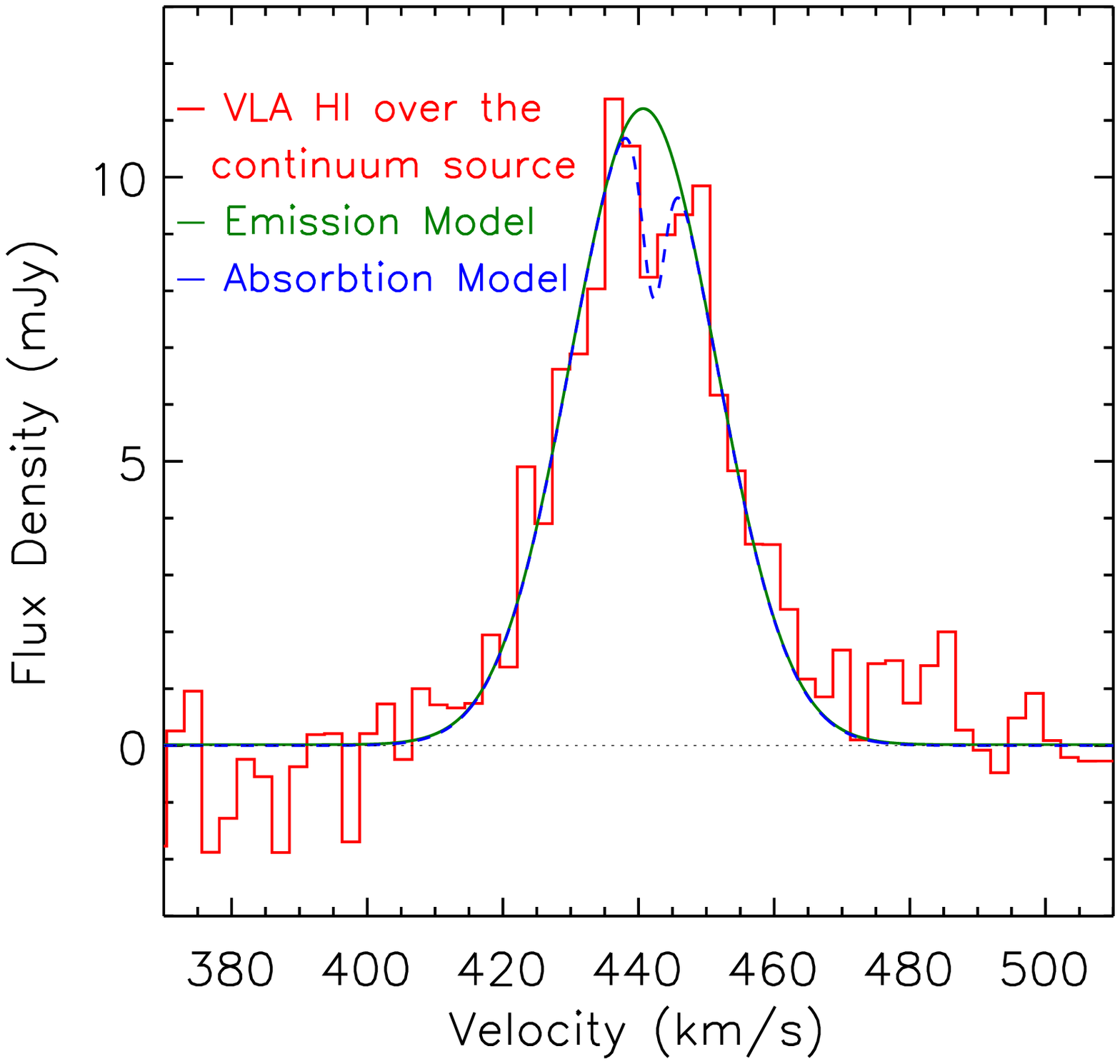}
\caption{ LEFT: A comparison plot showing the GBT HI spectrum from a region of size 9.1$^{\prime}$ (equivalent to the GBT beam at 21~cm) toward UGC~7408 in black and the VLA spectrum, obtained using natural weighting, encompassing the entire HI associated with UGC~7408 in blue. The good match between the two spectra proves that the VLA recovered almost 100\% of the \HI\ associated with this galaxy. The VLA \HI\ spectrum from the region marked by the black  rectangle (Figure~\ref{VLA_HI_image}, right panel) of size 40$^{\prime\prime} \times$50$^{\prime\prime}$ is shown in red in the comparison plot. The spectrum was obtained from data reduced using uniform weighting, which resulted in a better spatial resolution. RIGHT: A magnified version of the red spectrum showing two emission peaks and a dip in between at 442.2~\kms. Interestingly, the stronger peak at 436.3~\kms is toward the low-velocity end of the profile unlike the global \HI\ profile of the galaxy at that velocity. The green solid line and the blue dotted line represent Gaussian fits to the emission and the absorption profile respectively. The absorption is narrow with FWHM of 4.75~\kms corresponding to a kinetic temperature of $\le 493$~K. }
\label{VLA_HI_spec}
\end{figure*}

As we observed UGC 7408 with both the VLA and GBT, we were able to compare the HI emission from the galaxy that we obtained from each telescope. Figure~\ref{VLA_HI_spec} shows a good match between the spectra from the two telescopes. The VLA recovered integrated flux of 7.4~Jy~\kms (removing negative pixels) as compared to 8.1~Jy~\kms of GBT. This confirms that the VLA data recovered more than 91\%  of the \HI\ flux density detected by the GBT. The \HI\ spectral profile shows a deviation from a Gaussian profile at the peak of the emission and in the high velocity wing between 480 and 500~\kms.  The VLA spectrum obtained from a region of equal size as the synthesized beam in the uniform weighting case (the black rectangle in the left panel of Figure~\ref{VLA_HI_image}) over the quasar reveals two emission peaks with a dip at 442.2~km~s$^{-1}$. This could be a sign of absorption and is discussed in detail in section~\ref{sec:UGC7408}.

In the other two galaxies with \HI\ emission (J141629.25+372120.4 and J160659.13+271642.6), the expected \HI\ diameter estimated using the relationship shown in Figure~7 of \citep{swaters02} are 30.7 and 60.2~kpc, respectively.
 Thus, the sightlines, SL\#~22 and 23 are expected to pierce through the \HI\ distribution in these systems. 
 Although we detect dips in the GBT spectra, it is unclear if there is real absorption or are overlapping individual emission features mimicking absorption.
  Hence, we consider these two cases as nondetections. A detailed discussion of possible beam effects and how the GBT spectra in these two cases may be a result of the beam being filled-in by emission is presented in Section~\ref{sec:UGC7408}.

Besides the \HI\ emitting galaxies, we detected one unambiguous \HI\ absorption features at 21~cm in the foreground galaxy \galaxyname\ (SL\#10), a result we discussed in detail in \citet{bor10}. 
\galaxyname is a dwarf spiral galaxy of luminosity 0.048$~L_*$. The sightline pierces through the optical disk in this moderately star-forming galaxy. We estimated the kinetic temperature of the cloud to be $\le$ 283~K based on the absorption line FWHM of 3.6~\kms. Assuming the spin temperature to be same as the kinetic temperature, we estimated the column density of the absorber to be  $N$(\ion{H}{1})~$\leq~9.6~\times~10^{19}$~cm$^{-2}$. A detailed analysis of the properties of the absorber, the foreground galaxy, and its environment is presented in \citet{bor10}. 
We did not detect any \HI\ absorber in our GBT data from the three \HI\ emitting regions. 
Part of the problem may be the large GBT beam size, which can include a substantial portion of \HI\ emitting regions that may fill up the absorption. This possibility is also discussed in detail in section~\ref{sec:UGC7408}. Table~\ref{tbl-gbt_obs} presents the 3~$\sigma$ limiting optical depths for each of the sightlines along with the flux density sensitivity achieved. 
Assuming T$_{spin}$=100~K for the absorbing gas with $f_{cov}=1$ and a line-width of $\sim$5~\kms , the corresponding limiting column densities would range from 0.09 - 5.83 ~$\times ~ \rm 10^{20}~cm^{-2}$, with a median of ~1.46 $\times ~ \rm 10^{20}~cm^{-2}$.

\section{Discussion\label{sec:discussion}}

\subsection{Covering Fraction  \label{sec:gas_rare} }

\begin{figure*}
 \figurenum{5}
 \hspace{-0.5cm}
\includegraphics[trim = 0mm 0mm 50mm 0mm, clip,angle=0,scale=.56]{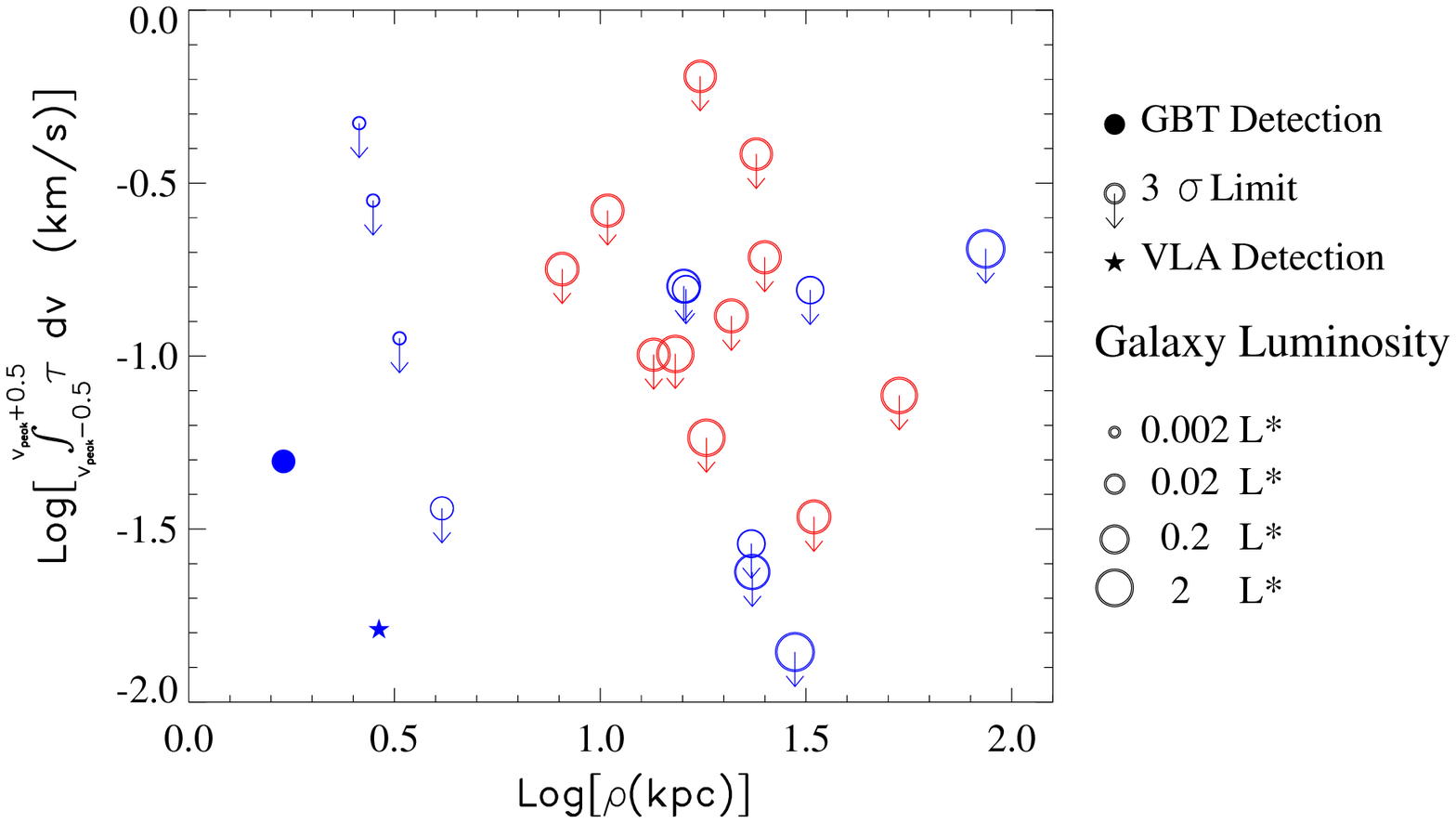}\includegraphics[angle=0,scale=.56]{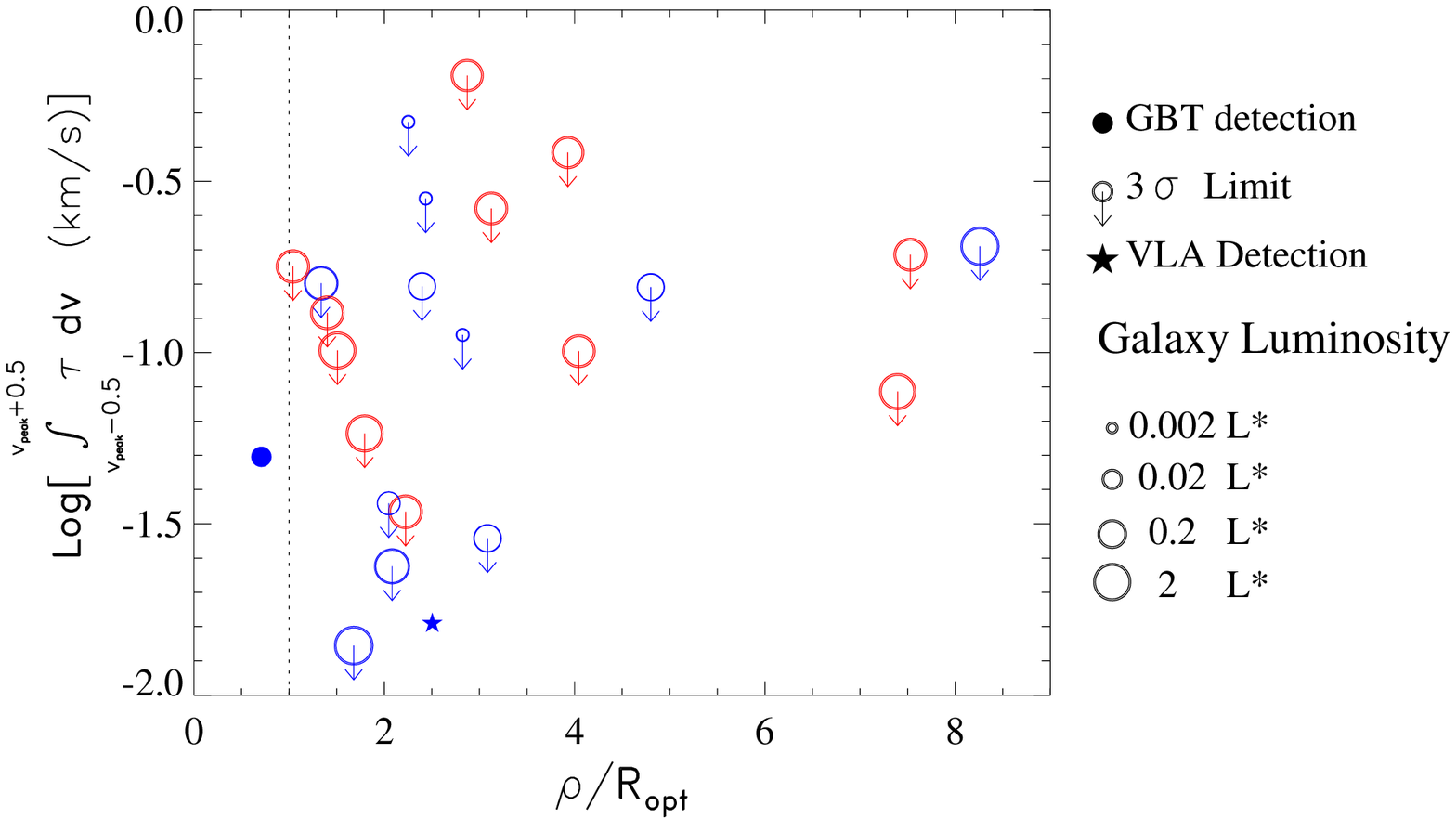}
\caption{LEFT: Impact parameter, $\rho$, versus the GBT integrated optical depth over channels corresponding to 1~\kms (i.e. 3~channels), $\int_{v_{peak}-0.5}^{v_{peak}+0.5}~\tau~\rm dv$, for our sample. The filled circles indicate 21~cm absorbers and open circles represents the 3~$\sigma$ limiting values. The colors show color classification of the galaxy as discussed in Section~\ref{sec:optical}. The peak optical depth per \kms for the absorber detected in the VLA data of UGC~7408 is shown as a filled star. RIGHT: Ratio of impact parameter to optical Petrosian r-band radius, $\rho/\rm R_{opt}$, versus the integrated optical depth. The dotted line represents $\rho/\rm R_{opt}=1$. The sightline probing \galaxyname\ is the only sightline to probe the optical disk. This is also our only unambiguous 21~cm absorption detection with the GBT. }
\label{rho_tau} 
\end{figure*}

The covering fraction of cold gas in galactic halos has been a matter of debate. While low-ionization gas in the \Lya and \ion{Mg}{2} transition have been detected in the halos of various galaxies, \HI\ 21~cm transitions have not been explored extensively.  
For instance, \citet{lanzetta95} found that a significant portion of the \Lya absorbers (1/3 to 2/3) were associated with normal, luminous galaxies with \Lya absorbing halos of size 160~$h^{-1}$kpc. They found an anti correlation between the strength of the absorption (equivalent width) and galaxy impact parameter, which was later confirmed by other authors \citep[e.g.][but see also Bowen et al. 2002; Wakker \& Savage 2009]{tripp98,chen98, chen01}. In this study, our aim is to estimate the frequency of occurrence of 21~cm absorbers in the halos and extended disks of galaxies. Figure~\ref{rho_tau} shows our findings in terms of two plots. The left plot shows the variation of measured optical depth as a function of impact parameter and the plot on the right shows the same quantity as a function of the ratio of impact parameter over the r-band Petrosian radius of the foreground galaxies. While the left plot shows the strength of the absorption or limiting strength versus the rest frame projected distance, the plot on the right shows the same quantity as a function of the optical size of the foreground galaxy. In addition, the optical colors of the foreground galaxies are denoted using respective colored symbols, and the luminosity with symbol-size, as described in the legend. Our single 21~cm absorber detection towards, \qsoname, is shown as a filled circle while all the other non-detections are shown with open circles with arrows pointing down. 
We do not include the limiting optical depth values for the \citet{gupta10} sample as their channel width is more than an order of magnitude higher than ours.  In addition, \citet{gupta10} presented their limiting optical depth for equivalent velocity resolution of 10~\kms. In contrast, we present our optical depth limits for a resolution of 1~\kms. The high resolution guarantees that such a study will detect absorbers similar to the one in \galaxyname\ or for \HI\ at a  temperature of $\sim$50~K.

The nondetection of cold \HI\ can be understood in terms of three possibilities: (1) the nonexistence of cold gas along the QSO line of sight, (2) a low covering fraction of the background QSO by the absorber or high spin temperature of the absorber, and/or (3) confusing effects due to the large beam size of the GBT (FWHM of 9$^{\prime}$.1). 
The possible existence of cold gas clouds with Damped \Lya (DLA) column density of 2$~\times~10^{20}~\rm cm^{-2}$ in the region surveyed can be estimated using the cross section implied by the $dN/dz$ of DLA systems \citep[see e.g., section 6.1 in][]{schaye07}, where the covering fraction of such clouds is described as 
\begin{equation}{\label{eq-fill_frac}}
f_{cov}^{DLA}~ \thickapprox~ 7.3 ~ \times ~ 10^9 ~ \Big( \frac{dN}{dz} \Big) ~\Big( \frac{n_g}{10^{-2}~\rm Mpc^{-3}}\Big)^{-1} ~\Big(\frac{r}{\rm pc}\Big)^{-2},
\end{equation}
where $dN/dz$ is the rate of incidence of DLA systems at $z=0$, $n_g$ is the comoving number density of galaxies affiliated with the absorbers, and $r$ is the radius of the absorber cross section. $dN/dz$ for  21~cm column densities comparable to those of DLAs in the nearby universe was computed by various authors from large \HI\ surveys like the {\it Arecibo Dual-Beam Survey} (ADBS) and {\it HI Parkes All Sky Survey }(HIPASS). \citet{rosenberg03} found $dN/dz = 5.3~ (\pm~ 1.3)~\times ~ 10^{-2}$ from the ADBS whereas \citet{ryan-weber03} and \citet{zwaan05} found $dN/dz = 5.8~ (\pm~ 0.6)~\times ~ 10^{-2}$ and $4.5~ (\pm~ 0.6)~\times ~ 10^{-2}$ respectively from the HIPASS. 
 For our calculation, we adopt $dN/dz = 5.3~\times ~ 10^{-2}$ from \citeauthor{rosenberg03}, which is also the median of the three values cited above. This value is consistent with the DLA incidence per unit redshift $n_{\rm DLA}(z)= (0.044 \pm 0.005)\times (1+z)^{1.27\pm0.11}$ for $0<z<5$ as described by  \citet{rao06} based on their HST survey of SDSS quasars with intervening \ion{Mg}{2}-\ion{Fe}{2} absorption systems. The number density of galaxies within a magnitude range can be estimated from the Schechter luminosity function \citep{schechter76}. For our analysis, we use the Schechter function fit of $\phi_* = 1.49~\times~10^{-2}~h^3~\rm Mpc^{-3}$, $M_*-5$log$_{10}h=-20.44$, and $\alpha=-1.05$, estimated by \citet{blanton03} for SDSS r-band data k-corrected to z=0.1. For the range of luminosities in our sample, $n_g \sim 2 \times \rm10^{-2}  Mpc^{-3}$. Based on the Equation~\ref{eq-fill_frac}, the $dN/dz$ statistics require $f_{cov}^{DLA}=1$ for cross section of 14~kpc or less. The covering fraction drops as the cross section of the galaxy increases.  For our sample, at a cross section radius of 55~kpc, $f_{cov}^{DLA}$,  drops to $\approx$0.08, and at 90~kpc it further drops to $\approx$0.02.

\begin{figure}
 \figurenum{6}
\centerline{\includegraphics[trim = 0mm 0mm 0mm 30mm, clip,angle=0,scale=.33]{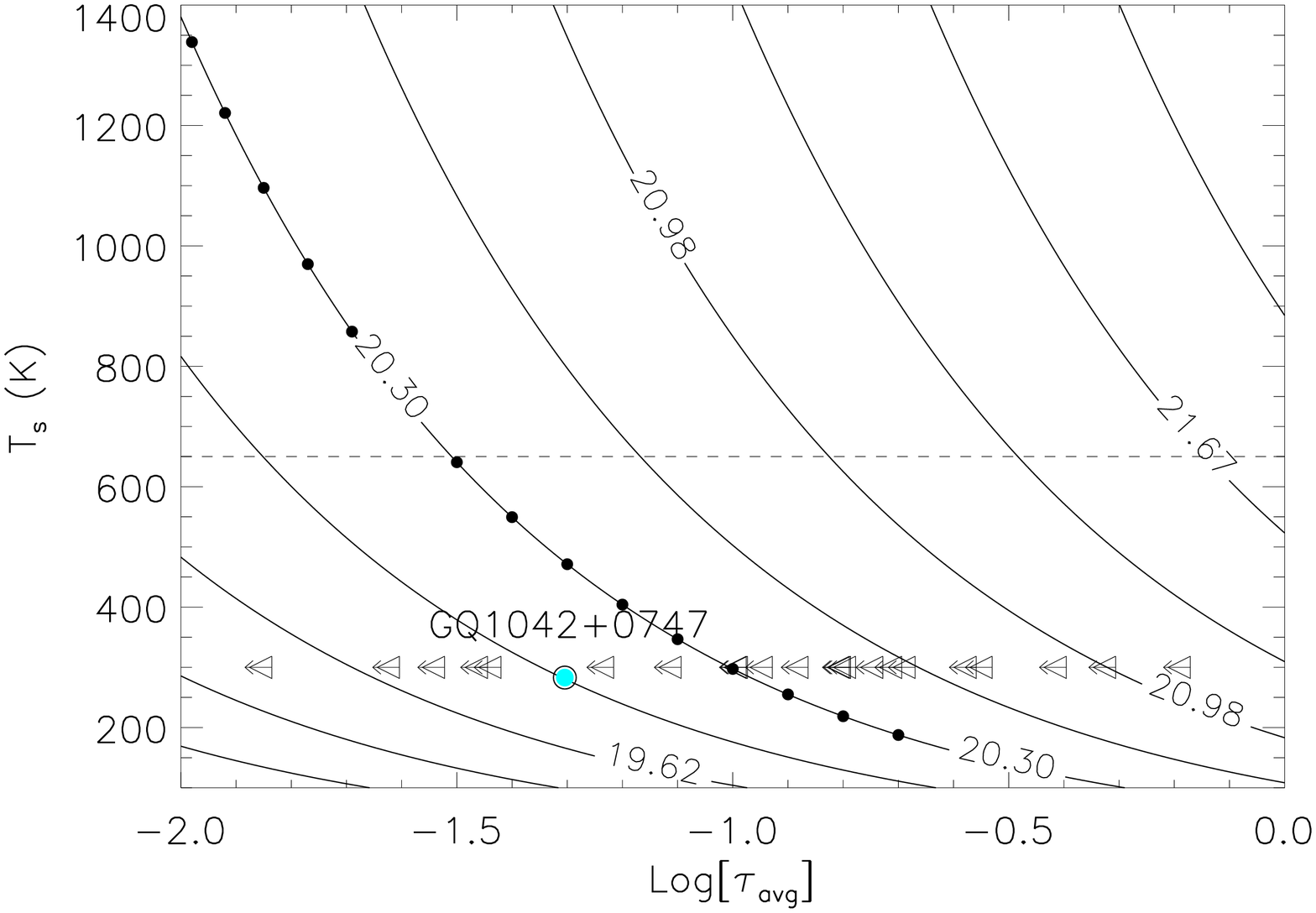}}
\caption{Variation of \HI\ column density as a function of average optical depth,$\tau_{\rm avg}$ and spin temperature, T$_s$ for absorbers with covering fraction, $f=1$. The contour levels mark column density in logarithmic scale. The contour at log~N(HI)= 20.3, which is indicated by a solid line with filled circles, represents the DLA limit of N(HI)=2$\times$10$^{20}~\rm cm^{-2}$. The horizontal dashed line represents the maximum T$_s$ detected in our Galaxy within the range of $\tau$ values shown in the plot \citep[see Fig 2a of][]{braun92}. The maximum T$_s$ for M31 is around 1000~K, although the majority of the data for both M31 and the Milky Way lie below 300~K. Our 21~cm absorber detected in the foreground galaxy \galaxyname, is plotted with a cyan circle. This is consistent with absorbers detected in the Milky Way. The GBT 3$\sigma$ limiting optical depths for the rest of our foreground galaxies that show no 21~cm absorption are plotted at T$_s=300~\rm K$ with arrow headed triangles and span a wide range of column densities. 
\label{N_HI_variation}}
\end{figure}

The presence of detectable 21~cm absorption not only depends on $f_{cov}^{DLA}$ and the column density of the gas but also on the spin temperature, T$_s$, as well as the fraction of the background quasar covered by the absorber. This fraction, which we will refer to as $f$ hereafter, which is different from $f_{cov}^{DLA}$. For an \HI\ absorption feature the column density can be estimated using the following equation from \citet{rohlfs00}:
\begin{equation}{\label{eq-col_den}}
N({\rm H~I}) = 1.823 \times 10^{18} \frac{T_{s}}{f}\int \tau _{21}(v) dv ~ cm^{-2},
\end{equation}
where $\tau_{21}(v)$ is the 21~cm optical depth as a function of velocity in km~s$^{-1}$. In cold clouds collisional excitation is the primary source of excitation for the hydrogen atom emitting the 21~cm line. Therefore, the above equation can be simplified by assume the spin temperature to be same as the kinetic temperature, T$_k$, which in turn can be related to the full-width at half maximum (FWHM) of the absorption feature by
 \begin{equation}{\label{eq-kin_temp}}
 T_{k}\le 21.855~(\Delta ~v)^{2}
 \end{equation}
where $\Delta v$ is the FWHM velocity in \kms. Therefore, Equation~\ref{eq-col_den} can be rewritten as 
\begin{equation}{\label{eq-col_T}}
N({\rm H~I}) = 0.390 \times 10^{18} \frac{T_{s}^{3/2}}{f} \tau _{avg} ~cm^{-2}
\end{equation}
where $\tau_{avg}$ represents the average optical depth such that $\tau_{avg}~\Delta~v~=~\int~\tau_{21}(v)~dv~$.
Figure~\ref{N_HI_variation} shows the variation of \HI\ column density as a function of optical depth, assuming $f=1$. The contours represent logarithms of \HI\  column density and are labeled accordingly. A column density of 2$~\times~10^{20}~\rm cm^{-2}$, corresponding to a DLA system, is shown with a solid line with filled circles overplotted and labeled 20.3. 
The filled cyan circle shows the absorber associated with \galaxyname. The dashed line represents the maximum spin temperature measured in the Milky Way \citep{braun92}, although the majority of their observations for the Milky Way and M31 have T$_s$ below 300~K. The absorber associated with \galaxyname\ is consistent with these observations. However, higher values of spin temperature have been commonly seen at higher redshifts and even in low-redshift absorbers associated with non-spiral galaxies \citep{lane98,chengalur00, kanekar01, kanekar05}. Our survey aims to detect cooler clouds with DLA equivalent column densities up to $1100~K$ (if $f=1$).
It is clear from Figure~\ref{N_HI_variation} that even for our best limiting optical depth value of 0.0139, if T$_s~>1100~K$ then the corresponding \HI\ column density would have to be well over the DLA limit to be detectable as a 21~cm absorber.
 On the other hand, we can be certain that a non-detection confirms that there are no cold clouds with T$_s < 100~K$ (assuming $f=1$) and column density equivalent to a DLA in our entire sample, with the exception of our noisiest sightline (SL\# 18 with $\tau_{avg}<0.64$).

\subsection{Cold gas in the Dwarf Galaxy UGC~7408  \label{sec:UGC7408}} 

The GBT spectrum towards 122106.854+454852.16 does not reveal any absorption from UGC~7408 despite the fact that there is \HI\ emission in the region in front of the quasar as seen in the VLA D-configuration image (see Figure~\ref{VLA_HI_image}). The \HI\ column density in the foreground region within the two lobes is between 1.44 and 2.88~$\times~ \rm 10^{20}~\rm cm^{-2}$ at the resolution of VLA D-configuration. In this section, we investigate the possible causes that may be responsible for the lack of absorption features in the GBT spectrum and what that may mean for the rest of our sample.
The nondetection of cold gas in absorption, despite the good sensitivity achieved, could be due to any of the three following underlying causes. For an \HI\ emitting galaxy such as UGC~7408, it is highly possible that the absorption is filled in by emission from the rest of the galaxy. This also extends to the other two \HI\ emitting foreground galaxies in our sample. The second possibility is that the \HI\ is mostly warm and does not absorb the background continuum. The third possibility is that the ISM in this galaxy is highly patchy and $N$(\ion{H}{1}) is much lower along the QSO sightlines than the average inferred at larger scales from the emission map.

Ideally, 21~cm absorption studies require high angular resolution to prevent absorption from being filled in by emission from the rest of the system. \citet{dickey00} suggests a resolution of $< \rm 10^{\prime\prime}$ for extragalactic sources. Unfortunately, the high sensitivity and spectral resolution achieved by the GBT comes at the cost of spatial resolution. 
On the other hand, our VLA D-configuration beam is much smaller than the GBT beam and this provides us with the opportunity to search for absorption from the region on the line of sight to the background quasar. The right panel of Figure~\ref{VLA_HI_spec} shows the VLA spectrum (in red) from the region marked by the black rectangle in Figure~\ref{VLA_HI_image}, which is also the size of beam for the uniform weighted data. The spectrum exhibits peaks at 436.3 and 449.3~\kms with a dip in between at 442.2~km~s$^{-1}$. 
The fact that the absorption dip sits right at the peak of the local emission suggests that the dip is an absorption feature. Furthermore, the ratio of the strength of the two peaks in the red spectrum is opposite to that of the global \HI\ profile of UGC~7408 at that velocity. This further suggests that we are spatially and kinematically resolving the underlying \HI\ emission. Even with the VLA D-configuration beam, we are still primarily detecting the emission from the foreground region.  For an \HI\ cloud and a background quasar contained within a single beam, the observed line temperature, $\rm T_L$, can be expressed as Equation~(12.22) of \citet{rohlfs00}
\begin{equation}
 {T_L}=(f_{cl}~T_s -f_0~f_C~T_C)(1-e^{-\tau} ) ,
\end{equation}
where, $f_{cl}$ and $f_C$ are beam filling factors of the \HI\ cloud and continuum source, respectively, $f_0$ is the fraction of the continuum source covered by the \HI\ cloud, and $T_C$ is the brightness temperature of the background source. From the FIRST image of background quasar, J122106.854+454852.16, we find that $f_C < 0.33$ and $T_C=25~K$ for the VLA in D-configuration, and thus we expect the emission to be significantly stronger than absorption even for very cold clouds ($T_s \sim~25~K$).

In order to get a proper estimate of the \HI\ column density associated with this dip, it is necessary to separate the emission profile from the absorption. By assuming the emission from the region covered by the VLA beam is a Gaussian, we fit the emission profile (excluding the dip). The green solid line in the right panel of Figure~\ref{VLA_HI_spec} shows the emission profile. Similarly the absorption profile is also fitted with a Gaussian profile yielding a FWHM=4.75~\kms centered at 442.2~\kms. The column density associated with this profile can be estimated using equation~\ref{eq-col_den}. If we integrate this profile and divide by the total background flux density of 67~mJy (background continuum flux density of 56~mJy $+$ background \HI\ emission flux density of 11~mJy), we get N(HI)= 3.8~$\times~ \rm 10^{19}$ ~$\frac{\rm T_s}{(100~K)}/f$ cm$^{-2}$. The kinetic temperature associated with this profile of FWHM =4.75~\kms\ is $\le 493~K$ and if $T_s = T_k$, then the column density associated with this absorber can be as high as $1.86~\times~\rm 10^{20} \rm ~cm^{-2}$ (assuming $f=1$).  The VLA D-configuration data may also suffer from emission filling-in the absorption as the beam encompasses a much larger region than the background continuum peaks. Higher resolution VLA A-configuration and/or VLBA data are needed to resolve the structures in the background quasar and to map the \HI\ only at the region in front of the continuum source. These observations will be crucial in understanding the small-scale structure of the ISM of UGC~7408. 

In the absence of higher resolution data, it cannot be proven that two peaks and the dip in the VLA spectra represents a signature of absorption. The feature may be part of the emission profile itself. If the feature were not an absorber, then it would mean that the \HI\ is at a higher temperature than that seen in the Cold Neutral Medium (CNM) of the Milky Way.  Spin temperatures of around 1000~K or higher have been seen at low-redshifts ($z< 0.5$) in non-spiral systems \citep{kanekar01}. \citet{dickey00} found that in the SMC the fraction of cool-phase \HI\ is much lower than the overall \HI\ abundance. The ISM properties of dwarf galaxies are different from bigger galaxies like the Milky Way and this may be the reason for the low cold-\HI\ abundances in their ISM. Dwarf galaxies have lower abundances of metals and thus, cooling from fine-structure lines (especially from carbon and oxygen) in these galaxies is much less predominant than brighter galaxies (see Dickey et al. 2000 for details). 
In addition, the typical dust to gas ratios in dwarf galaxies are much lower than in the Milky Way and thus dwarf galaxies are inefficient in confining the UV-ionizing radiation to individual \ion{H}{2} regions. Therefore, UV radiation can effectively heat up the cooler gas in dwarf galaxies to a larger radii outside the stellar disks. Hence, due to increased heating and reduced cooling, it is very likely that most of the \HI\ in dwarf galaxies is at a much higher temperature.

Another possibility for the nondetection of \HI\ absorption is the lack of \HI in the foreground region. This may be a result of a clumpy ISM, which when observed with VLA D-configuration resolution get smoothed out. 
\citet{puche92} found holes in the \HI\ distribution of Holmberg II, a much-studied star-forming dwarf galaxy, with diameters ranging from $\sim$ 100 to 1700~pc.
Similarly, \citet{begum08} also found gaps and/or holes in the \HI\  distribution for dwarf galaxies such as DDO43. Maps from the WHISP survey also show that galaxies have similar patches of low intensity \ion{H}{1} emission even in the inner regions. In the case of UGC~7408, since we have three sightlines through the ISM at impact parameters 2.6, 2.8, and 3.3~kpc, according to the last possibility, the ISM needs to be extremely clumpy. Also it requires the clumps to have very high column densities so that the average column density in emission when smoothed to VLA D-configuration resolution is around 2~$\times~ \rm 10^{20}~\rm cm^{-2}$. 
In conclusion, we emphasize the need for high spatial resolution VLA A-configuration imaging. These observations will not only provide information on the clumpiness of the ISM in dwarf galaxies but also will shed light on the spin temperature of the \HI\ in such systems. Furthermore, the background QSO is a UV bright source and can be observed with the {\it Cosmic Origin Spectrograph} (COS) on the {\it Hubble Space Telescope} (HST). The spectroscopic UV data would provide complementary information which will allow us to learn more about the multiphase ISM including the warmer phases of neutral gas in UGC~7408.

\subsection{Connection of 21~cm Absorbers to ISM of Galaxies \label{sec:absorber_ISM}}

Our only unambiguous GBT detection comes from the foreground galaxy \galaxyname, which is intercepted by a QSO sightline that has the smallest impact parameter of all the QSO-galaxy pairs in our sample, and which pierces the stellar disk of the galaxy. As discussed in \citet{bor10}, the absorber is very narrow and has a column density of $N$(\ion{H}{1})~$\leq~9.6~\times~10^{19}$~cm$^{-2}$,  i.e., lower than a DLA. It is evident from our sample that we do not find cold clouds with column densities of the order of DLAs or more, associated with the halos of galaxies ubiquitously. The second possible absorber, associated with UGC~7408, probably arises in the bound ISM of the dwarf galaxy.

Other studies have found 21~cm absorbers and some of them have been associated with tidally stripped matter. For example, \citet{stocke91} found a 21~cm \HI\ absorber associated with a tidal feature of NGC~3067.
Cool gas associated with galactic outflows has been observed in the halos of galaxies probed by various metal-line transitions. \citet{veilleux05} observed such outflows in nearby starburst galaxies and \citet{bc03} and \citet{keeney06} found indications of a bound outflow in the central region of the Milky Way. However, the nondetection of cold gas in the halos of galaxies in our sample suggests that either the amounts of \HI\ in these systems are not high enough or $T_s > 1100$ for the outflowing material. Since most of the measurements of outflowing material come from metal-line transitions, the temperature of the gas may be well above our limit to detect 21~cm \HI\ absorption even for our best-case sensitivity scenario. In addition, the radio background quasars are often extended and thus a clumpy structure of out-flowing material can have a relatively low covering fraction, $f$, thus significantly increasing the column density associated with our limiting optical depth.
 
The same argument can also be extended to accreting material or cold clouds condensing out of the halos. The above conclusions hold irrespective of the metallicity of the cold clouds, and therefore provide information about accreting and outflowing material with the same accuracy. Out of 15 foreground galaxies in our sample, three of them have masses $< 10^{10.3}~\rm M_{\odot}$ with no detectable \HI\ emission, and sightlines toward the background quasars through the halo of the galaxies. However we did not find any 21~cm absorber in their halos. This is not surprising as ``cold'' accretion could easily have spin temperatures that makes it difficult to detect in 21~cm absorption. We also do not find ubiquitous 21~cm absorption in the halos of luminous galaxies unlike \Lya studies.

\section{Summary \label{sec:conclusion}}

The process of gas flow into galaxies and subsequent condensation into cold gas has been a topic of various observational and numerical studies. In this paper, we investigated the presence of cold gas in the halos and extended disks of low-redshift galaxies. We find the occurrence of \HI\ 21~cm absorption is rare (1/12) as we move outwards from the optical disk of the galaxies for our sample. We exclude the three, \ion{H}{1}-emitting foreground galaxies as nondetections as any absorption along the background lines of sight may have been filled in by emission from \HI\ in the ISM of the galaxy itself. The only case where we unambiguously discovered atomic gas in absorption was in the stellar disk of the dwarf galaxy \galaxyname\ at an impact parameter of 1.7~kpc from the QSO sightline \qsoname. For systems with $\rho < 7.5$~kpc, which was reported as the median impact parameter for \HI\ systems corresponding to DLA column densities by \citet{zwaan05}, we detected \HI\ in one out of two galaxies. Our detection rates for sightlines with $\rho<15$~kpc do not match the results from \citet{gupta10}. This may be partly because of the inclusion of pairs selected due to the presence of Ca II and Na I absorption \citep[sightlines from][]{carilli92} in the \citet{gupta10} sample.

On the other hand, we detected \HI\ in emission in three foreground galaxies including the dwarf galaxy UGC~7408. The \HI\ masses range from 0.9 to 81.3~$\times~ \rm 10^{8}~ \rm M_{\odot}$.
 The VLA \HI\ imaging of UGC~7408 shows that its \HI\ envelope extends to 2-3 times the optical size of the galaxy in the form of an ellipsoid extending in the southeast direction. Interestingly, the background continuum sources (SL\#~14, 13, and 12) at impact parameters of 2.6, 2.8, and 3.3~kpc from the optical center of the galaxy is well within the \HI\ distribution seen in VLA imaging. We did find a dip in the \HI\ emission at the position of the background quasar. The dip is centered at 442.2~\kms and has a FWHM of 4.75~\kms corresponding to a kinetic temperature of $\le 493~$K. Assuming the spin temperature to be same as the kinetic temperature, we estimate the column density associated with this dip to be 1.86~$\times \rm 10^{20}~ \rm cm^{-2}$. However, higher resolution imaging is required to confirm this detection, because the dip may be a result of the emission profile from the foreground ISM of UGC~7408 and not an absorption signature. This may suggest that the \HI\ seen in the VLA imaging may have too high a temperature to detect 21~cm absorption or the \HI\ distribution in this galaxy is highly patchy with no \HI\ structures at the position of the background continuum source.

Our study was designed to be most sensitive to cold gas with temperatures observed in the CNM of the Milky Way. However in a few cases our sensitivities are high enough to look for DLAs at temperatures above 500~K. For instance, along the sightline where we achieved the best 3~$\sigma$ limiting optical depth of 0.01, we were sensitive to DLA systems even at $>$1000~K. On the other hand, for low temperature systems, this means that we could detect absorbers with column densities as low as  5.6~$\times$~10$^{18}~\rm cm^{-2}$ if T$_s~= \rm 100~K$. Therefore, we can conclude that in most cases, there is no \HI\ with DLA column densities at temperatures similar to that of CNM in Milky Way in the halos and extended disks of galaxies. Although our data are insufficient to confirm the inflow and/or outflow of neutral gas through the galactic halos, our nondetection of \HI\ in the halos does hint that the process of condensation of warmer gas into \HI\ may be occurring at regions closer to the optical disk of galaxies. 
Our present understanding of the cold gas distribution in the halos and extended disks of galaxies is quite limited. To extend our knowledge, high-resolution \HI\ imaging of galaxy-quasar pairs as well as complementary information about warmer phases of gas through UV observations is crucial. In the future, the Expanded VLA (EVLA) with broad bandwidth and the COS with high sensitivity will be able to probe systems through observations that would yield information on the physical nature and distribution of various gaseous species in the halos and extended disks of galaxies.\\

\acknowledgements
 This paper has benefited from technical suggestions from Yuxi Chen, Yicheng Gao, and Jason Prochaska.
The authors are grateful to the observatory staff at the GBT and the VLA who made these observations possible. 
 SB is grateful for a Student Observing Support award (GSSP08-0024) from the National Radio Astronomy Observatory that made this work possible.
 This research was also supported by NASA ADP Grant NNX08AJ44G. 
 DVB is funded through NASA LTSA grant NNG05GE26G.
 Funding for the SDSS and SDSS-II has been provided by the Alfred P. Sloan Foundation, the Participating Institutions, the National Science Foundation, the U.S. Department of Energy, the National Aeronautics and Space Administration, the Japanese Monbukagakusho, the Max Planck Society, and the Higher Education Funding Council for England. 
 The SDSS Web Site is http://www.sdss.org/. 
 The SDSS is managed by the Astrophysical Research Consortium for the Participating Institutions. 
 The Participating Institutions are the American Museum of Natural History, Astrophysical Institute Potsdam, University of Basel, University of Cambridge, Case Western Reserve University, University of Chicago, Drexel University, Fermilab, the Institute for Advanced Study, the Japan Participation Group, Johns Hopkins University, the Joint Institute for Nuclear Astrophysics, the Kavli Institute for Particle Astrophysics and Cosmology, the Korean Scientist Group, the Chinese Academy of Sciences (LAMOST), Los Alamos National Laboratory, the Max-Planck-Institute for Astronomy (MPIA), the Max-Planck-Institute for Astrophysics (MPA), New Mexico State University, Ohio State University, University of Pittsburgh, University of Portsmouth, Princeton University, the United States Naval Observatory, and the University of Washington.

{\it Facilities:}  \facility{ARC ()}, \facility{GBT ()}, \facility{Sloan ()}, \facility{VLA ()}

\end{document}